\documentclass[12pt]{article}
\usepackage{amssymb}
\usepackage{amsmath}
\usepackage{amstext}
\usepackage{graphicx,epsfig}
\usepackage{epsfig}
\usepackage{verbatim} 
\usepackage{fancyhdr}
\usepackage{fancybox}
\usepackage{color}
\usepackage{ulem,bbold}
\usepackage{enumitem}
\usepackage{subfigure}
\usepackage{bbm}
\usepackage{parskip}
\usepackage{cite}

\newcommand{\Comment}[1]{{}}
\definecolor{MyDarkBlue}{rgb}{0.15,0.15,0.45}
\usepackage[linktocpage=true]{hyperref}
\hypersetup{
colorlinks=true,
citecolor=MyDarkBlue,
linkcolor=MyDarkBlue,
urlcolor=MyDarkBlue,
}

\newcommand{\be}{\begin{equation}}
\newcommand{\ee}{\end{equation}}
\newcommand{\bea}{\begin{eqnarray}}
\newcommand{\eea}{\end{eqnarray}}
\newcommand{\beas}{\begin{eqnarray*}}
\newcommand{\eeas}{\end{eqnarray*}}
\newcommand{\nn}{\nonumber}
\newcommand{\la}{\langle}
\newcommand{\ra}{\rangle}

\newcommand{\half}{\frac{1}{2}}

\numberwithin{equation}{section}

\usepackage{genyoungtabtikz}
\Yboxdim{13pt} 
\Ylinethick{.8pt}

\begin{document}

%\maketitle

\begin{center}
{\Large \bf{Shift Symmetries and AdS/CFT\\ }}
\vspace{.2cm}
\end{center}

\vspace{1.5truecm}
\thispagestyle{empty}
\centerline{\Large Erin Blauvelt,${}^{\rm a,}$\footnote{\href{mailto:e.k.blauvelt@gmail.com} {\texttt{e.k.blauvelt@gmail.com}}} 
Laura Engelbrecht,${}^{\rm b,}$\footnote{\href{mailto:lxj154@case.edu}{\texttt{ljohnson@phys.ethz.ch}}} 
 Kurt Hinterbichler,${}^{\rm c,}$\footnote{\href{mailto:kurt.hinterbichler@case.edu} {\texttt{kurt.hinterbichler@case.edu}}}  }

\vspace{.5cm}
 
 \centerline{{\it ${}^{\rm a}$Department of Physics,}}
 \centerline{{\it Lehigh University, Bethlehem, PA, 18018}} 
  \vspace{.25cm}
 
 \centerline{{\it ${}^{\rm b}$Institute for Theoretical Physics,}}
 \centerline{{\it ETH Zürich, Wolfgang-Pauli-Strasse 27, 8093, Zürich, Switzerland
}} 
   \vspace{.25cm}
 
   \centerline{{\it ${}^{\rm c}$CERCA, Department of Physics,}}
 \centerline{{\it Case Western Reserve University, 10900 Euclid Ave, Cleveland, OH 44106}} 

 \vspace{1cm}

\begin{abstract} 

Massive fields on anti-de Sitter (AdS) space enjoy galileon-like shift symmetries at particular values of their masses.  We explore how these shift symmetries are realized through the boundary conformal field theory (CFT), at the level of the 2-point functions.  
In the alternate quantization scheme in which the dual conformal field gets the smaller $\Delta_-$ conformal dimension, the shift symmetry is realized as a gauge symmetry in the dual CFT, so that only shift invariant operators are true conformal primary fields. In the standard quantization scheme the shift symmetry acts on the source, leading to Ward identities that take the form of integral constraints.

\end{abstract}

\newpage

\thispagestyle{empty}
\tableofcontents

\setcounter{page}{1}
\setcounter{footnote}{0}

\parskip=5pt
\normalsize

\section{Introduction}

The Anti-de Sitter (AdS) space/conformal field theory (CFT) correspondence \cite{Maldacena:1997re} allows us to view physics on AdS as an emergent phenomenon out of the quantum mechanical degrees of freedom of a boundary CFT.  AdS/CFT has gone far beyond its initial formulation, and over the last 20 years the relationship has provided insight into a wide range of physical systems with significant inherent challenges. In particular, this duality provides an avenue to probe strongly coupled quantum
systems in nature that are far from equilibrium, ranging from cosmology to condensed matter \cite{Hartnoll:2016apf,Hartnoll:2009sz,Baggioli:2019rrs,Antonini:2019qkt}. 

The symmetries of a system hold a wealth of information about the system's behavior. Shift symmetries, in particular, generally arise in systems which have undergone spontaneous symmetry breaking. Spontaneous symmetry breaking commonly plays a role in holographic applications to condensed matter systems \cite{Ammon:2019wci,Cai:2015cya}, yet, the role of shift symmetries in holography remains largely unexplored (see however \cite{Marzolla:2017ddu,Marzolla:2017sfd}).

The possible AdS covariant shift symmetries acting on massive symmetric tensor fields on AdS were catalogued in \cite{Bonifacio:2018zex}. These symmetries generalize the extended shift symmetries of the flat space galileon \cite{Luty:2003vm,Nicolis:2008in} and special galileon \cite{Cheung:2014dqa,Hinterbichler:2015pqa,Cheung:2016drk,Novotny:2016jkh} to AdS and to higher spins.  (See the introduction of \cite{Bonifacio:2018zex} for more  background and motivation for these symmetries.) 

A generic massive field of spin $s$ on AdS$_D$, with AdS radius $L$, is carried by a fully symmetric tensor field $\Phi_{\mu_1\cdots\mu_s}$.  The equations of motion are the Klein-Gordon equation,
\be \begin{cases} \left(\nabla^2 -m^2\right)\Phi=0\,\, ,\ \ \ \ \ \ \ \ \ \ \ \ \ \ \ \ \ \ \ \ \ \ \ \ \ \ \ \ \ \ \ \ \ \ \ \ \ \ \ \ \ \ \ \ \ \ \ \ \ \ \ \ \ \ \ \ \ \ \ s=0\,, \\
\left(\nabla^2+{1\over L^2}\left[s+D-2-(s-1)(s+D-4)\right]-m^2\right)\Phi_{\mu_1\cdots\mu_s} = 0\ ,\  \ s\geq 1\,, \label{spinsonsheeqede}
\end{cases} \ee
along with transversality in all the indices, and tracelessness in all the indices.

As detailed in \cite{Bonifacio:2018zex}, the shift symmetries are present for certain masses, $m_k$, where $k$ is called the level of the shift symmetry,
\be \begin{cases} m_{k}^2L^2=k(k+D-1),\ \ \ \ \ \ \ \ \ \ \ \ \ \ \ \ s=0\, , \\
m^2_{k}L^2= (k+2) (k+D-3+2 s) ,\ \  s\geq 1\, ,
\end{cases} \ \ \  \ \ \ \ k=0,1,2,\ldots \,. \label{symtenpmmassee}
\ee
The explicit form of the shift symmetry is 
\be \label{eq:tensorshift}
\delta \Phi_{\mu_1\cdots\mu_s}= S_{A_1\cdots A_{s+k},B_1\cdots B_s}X^{A_1}\cdots X^{A_{s+k}} {\partial X^{B_1}\over \partial x^{\mu_1}}\cdots {\partial X^{B_s}\over \partial x^{\mu_s}} \, ,
\ee
where the $X^A(x)$ are embedding coordinates of the AdS$_D$ spacetime into an auxiliary flat ambient spacetime of dimension $D+1$, and the tensor $S_{A_1\cdots A_{s+k},B_1\cdots B_s}$ is a constant traceless tensor whose indices have the symmetries of a two row Young tableau,
\be 
{S_{A_1\cdots A_{s+k},B_1\cdots B_s}\in ~\raisebox{1.15ex}{\gyoung(_5{s+k},_3s)} \, } .
\label{repKexpre}
\ee

As discussed in \cite{Bonifacio:2018zex}, the shift symmetric fields can be constructed as the decoupled longitudinal mode of a massive field as it approaches a partially massless (PM) value.  The PM fields for spin $s\geq 1$ occur at mass values labeled by an integer, $t$, called the depth,
\be
\bar{m}^2_{t} =-{1\over L^2} (s-t-1)(s+t+D-4)\,,\ \ \ t=0,1,\ldots,s-1\,.
\label{pmpoints}
\ee
The massless field is the special case $t = s-1$.  A depth-$t$ PM field has a gauge invariance where the gauge parameter is a rank-$t$ totally symmetric tensor, which on-shell looks like
\be
 \delta \Phi_{\mu_1\cdots\mu_s} = \nabla_{(\mu_{t+1}}\nabla_{\mu_{t+2}}\cdots\nabla_{\mu_{s}}\xi_{\mu_{1}\cdots\mu_{t})_T}\,. \label{pmintogt}
\ee

The $k$-th shift symmetric spin $s$ field can be constructed as the decoupled longitudinal mode of a massive spin $s+k+1$ field as its mass approaches the depth $s$ PM value. The shift symmetric field has a symmetric and traceless ``field strength'' tensor $F_{\mu_1\ldots \mu_{s+k+1}}$ with the same indices as this partially massless parent field.  It is constructed from the $(k+1)$-th symmetrized traceless derivative, in a manner patterned after the PM gauge transformation, 
\be F_{\mu_1\ldots \mu_{s+k+1}}= \nabla_{(\mu_{s+1}} \cdots \nabla_{\mu_{s+k+1}}\Phi_{\mu_1\ldots \mu_s)_T}\,  .\label{shifsymfeste} \ee
This field strength is invariant under \eqref{eq:tensorshift} and gives the basic shift-invariant operator in the theory. 

For a generic massive spin $s$ field of mass $m$, the dual operator on the boundary in the $d=D-1$ dimensional CFT is also of spin $s$, and has a scaling dimension $\Delta$ related to the AdS mass by
\be \begin{cases} m^2L^2=\Delta(\Delta-d)\,\, ,\ \ \ \ \ \ \ \ \ \ \ \ \ \ \ \ \ \ \ \ \ \ \ \ \ s=0\,, \\
m^2L^2=(\Delta+s-2)(\Delta-s-d+2)\ ,\  \ s\geq 1\,.
\end{cases} \ee
The greater and lesser roots of this equation are given by
\be \begin{cases} \Delta_\pm={d\over 2}\pm \sqrt{{d^2\over 4}+m^2L^2}\,\, ,\ \ \ \ \ \ \ \ \ \ \ \ s=0\,, \\
\Delta_\pm= {d\over 2}\pm\sqrt{{\left(d+2(s-2)\right)^2\over 4}+m^2L^2}\ ,\  \ s\geq 1\,.
\end{cases} \ee
There are two ways of setting AdS invariant boundary conditions when quantizing the AdS field, which lead to two different CFTs: $\Delta_+$ gives the operator dimensions in the dual CFT obtained by standard quantization, and $\Delta_-$ gives the value obtained by alternate quantization \cite{Klebanov:1999tb}. 
For the shift symmetric fields, we have
\be \Delta_+=d+s+k\, ,\ \ \ \Delta_-=-s-k\, ,\ \ \label{confordimssfe}\ee 
and for the PM fields, we have
\be \Delta_+=t+d-1\, ,\ \ \ \Delta_-=1-t \, \ . \label{PMconfordimssfe}\ee

In the Poincar\'e patch of AdS the metric reads
\be ds^2={L^2\over z^2}\left(dz^2+\delta_{ij}dx^i dx^j\right),\label{adspoincaremete}\ee
where $z\in (0,\infty)$ with  $z=0$ the boundary and $z=\infty$ the deep interior, and $\delta_{ij}$ is the flat boundary metric.
In these coordinates, the boundary components of a spin-$s$ field of generic mass have a near boundary Fefferman-Graham expansion \cite{AST_1985__S131__95_0} that contains two leading fall-off behaviors, 
\be \Phi_{i_1\cdots i_s}(x,z)= z^{\Delta_- -s}\left[\phi_{(0)i_1\cdots i_s}(x)+\cdots\right]+ z^{\Delta_+ -s}\left[\psi_{(0)i_1\cdots i_s}(x)+\cdots\right]. \ee
In standard quantization, $\phi_{(0)i_1\cdots i_s}$ is proportional to the source in the dual CFT's generating functional, whereas in alternate quantization $\psi_{(0)i_1\cdots i_s}$ is proportional to the source.  We will see that the shift symmetry acts only on $\phi_{(0)i_1\cdots i_s}$, in a manner that shifts it by a generalized conformal Killing tensor.  When $\phi_{(0)i_1\cdots i_s}$ is fixed to be the source, the shift symmetry then acts on the source. In this case, the correlation functions have the canonical structure for conformal primary operators and the symmetry manifests itself in the dual CFT as integral Ward identities. However $\psi_{(0)i_1\cdots i_s}$ is not affected by the shift symmetry, and when $\psi_{(0)i_1\cdots i_s}$ is fixed to be the source we will instead see a gauging of the shift symmetry directly in the dual CFT.

This gauging of the shift symmetry in alternate quantization shows up as logarithms in the correlators in position space, which break conformal symmetry. These are analogous to the logarithms that appear in the scalar field 2-point function for a free massless scalar in $d=2$.  
The logarithms lead us to interpret the shift symmetry as being gauged, so that only operators invariant under the shift symmetry are true operators in the theory.  The basic invariant operator is the boundary value of the shift symmetric field strength tensor \eqref{shifsymfeste} made from $k+1$ symmetrized derivatives of the field.  Taking the derivatives to form the field strength, the logs in the position space correlator of the original spin $s$ weight $-k-s$ field go away, and the correlator takes the proper conformally invariant form for a field of spin $s+k+1$ and dimension $1-s$.

This all goes through straightforwardly when $d$ is odd.  When $d$ is even, there are logarithmic terms in the near-boundary expansion of the shift symmetric field.  These give additional difficulties in interpreting the alternate quantization, which we discuss but do not fully resolve. 

\vspace{.15cm}
\noindent

{\bf Conventions:}
We use the mostly plus metric signature. 
The spacetime dimension is denoted $D$, with indices $\mu,\nu,\ldots$.   The dual CFT dimension is $d \equiv D-1$, with indices $i,j,\ldots$.  We restrict to $d\geq 3$ to avoid subtleties that happen in lower dimensions.  The AdS$_D$ radius is denoted by $L$.  The Ricci scalar is then $R=-{D(D-1)/ L^2}<0$.  $X^A(x)$ denotes the embedding of AdS$_D$ into flat spacetime of dimension $D+1$, with indices $A,B,\ldots$, with conventions as given in appendix A of \cite{Bonifacio:2018zex}.   
Tensors are symmetrized and antisymmetrized with unit weight, {\it e.g.} $t_{[\mu\nu]}=\half \left(t_{\mu\nu}-t_{\nu\mu}\right)$, and $(\cdots)_T$ denotes the symmetric fully traceless part of the enclosed indices.  Young tableaux are in the manifestly symmetric convention. Our Fourier transform convention for 2 point functions is $\la \phi(p)\phi(-p)\ra=\int d^d x e^{i p\cdot  x}\la \phi(x)\phi(0)\ra \,.$

\section{Shift Symmetries and CFT 2-point Functions\label{2ptsec}}

We start from a purely CFT point of view, by studying the 2-point functions of conformal primary fields at the values of the conformal dimensions \eqref{confordimssfe} corresponding to the shift symmetric fields.  We will see that the shift symmetry manifests itself at the $\Delta_-$ value.  

In a unitary CFT theory, conformal primaries must satisfy the bound
\cite{Mack:1975je,Jantzen1977,Minwalla:1997ka},
\be \Delta\geq  \begin{cases}  {d\over 2}-1 , & s=0\,, \\ s+d-2, & s\geq 1\, .\end{cases}\label{unitboundde}\ee
We can see that the $\Delta_+$ values in \eqref{confordimssfe} all lie above the unitarity bound, and the $\Delta_-$ values all lie below it, so we will be interested in behavior that occurs in the non-unitary regime.

In position space, the 2-point function of two identical primaries is completely fixed by conformal symmetry up to an overall constant, which we remove by rescaling the operator.  
The 2-point function of such a canonically normalized spin-$s$ primary $\phi_{ i_1\cdots i_s}(x)$ of weight $\Delta$ is fixed to be \cite{Osborn:1993cr}
\be \la \phi_{i_1\cdots i_s}(x)\phi^{ j_1\cdots j_s}(0)\ra={1\over x^{2\Delta}}I_{( i_1}^{( j_1}\cdots I_{ i_s)_T}^{ j_s)_T},\ \ \ I_{ i j}\equiv \delta_{ i j}-2{x_ i x_ j\over x^2}.\label{gencoonfprimarsse}\ee

\subsection{Scalars\label{sec1scalarsubsec}}

We start with the simplest case of scalar primary operators, $s=0$.  The position space 2-point correlator at separated points for a scalar of weight $\Delta$ is
\be \la \phi(x)\phi(0)\ra={1\over x^{2\Delta}}.\label{position2pse}\ee
Directly Fourier transforming using the methods of \cite{Freedman:1991tk}, we get
\be \la \phi(p)\phi(-p)\ra= \frac{\pi ^{d/2}   }{2^{2 \Delta-d }}   \frac{  \Gamma \left({d\over 2}- \Delta \right) }{\Gamma (\Delta )}  p^{2 \Delta -d} .\label{pspace2pse} \ee

The expression \eqref{pspace2pse} has poles at the values of $\Delta$ for which
\be \nu\equiv \Delta-d/2=0,1,2,3,\ldots\ \ ,\label{scalarpolevaluese}\ee
which occur for the $\Delta_+$ shift symmetric values in even $d$.  The values \eqref{scalarpolevaluese} are where the momentum dependence $\sim p^{2 \Delta -d}$ in \eqref{pspace2pse} becomes local.  In these cases \eqref{pspace2pse} is not the desired Fourier transform of \eqref{position2pse}.  The desired Fourier transform shows up as the log terms in the finite part of the Laurent expansion of \eqref{pspace2pse} about the pole: taking $\Delta\rightarrow {d\over 2}+\nu+\epsilon$ and expanding in $\epsilon$ we find a local divergence and logarithmic finite part,
\be \la \phi(p)\phi(-p)\ra \rightarrow  
   \frac{\pi ^{d/2} (-1)^{\nu +1}}{\nu !  4^{\nu } \Gamma \left(\frac{d}{2}+\nu \right)} \left({1\over \epsilon}  p^{2 \nu }+p^{2 \nu } \log (p^2)\right)+\cdots  
  \label{2scalarlogsolutionee}
   \ee
Adding the local conformally  invariant counterterm $\sim p^{2 \nu}$ cancels the divergence and can be used to adjust the scale of the log.  The resulting finite correlator is the Fourier transform of \eqref{position2pse} with its short distances singularities appropriately subtracted off (see e.g. \cite{Freedman:1991tk,Petkou:1999fv,Bianchi:2001kw,Skenderis:2002wp}).  These momentum space logs do not seem to have anything to do with the shift symmetry of the corresponding AdS field; indeed they occur in odd $d$ for half integer values of $\Delta$, whose bulk fields do not carry the shift symmetry. 

The expression \eqref{pspace2pse} has zeros at values of $\Delta$ for which
\be k\equiv -\Delta=0,1,2,3,.\ldots\ \ .\label{scalarzeoersvalueee}\ee  
These are precisely the $\Delta_- $ shift symmetric values.  At these values the position space correlator \eqref{position2pse} becomes a polynomial in $x^2$.  The Fourier transform of such a polynomial is a local contact term in momentum space.  Since the Fourier transform method leading to \eqref{pspace2pse} is blind to such local contact terms, these polynomial terms become trivial in momentum space, and hence we get zeros in \eqref{pspace2pse}.   The values \eqref{scalarzeoersvalueee} are also special in that the conformal algebra dictates that the primary operator satisfies a shortening condition where the $(k+1)$-th symmetrized traceless derivative vanishes \cite{Penedones:2015aga}
\be \partial_{(i_1}\cdots\partial_{i_{k+1})_T}\phi=0.\ee
This is clearly satisfied by a correlator which is polynomial in $x^2$.

These polynomial correlators are not what we typically want in a local field theory, where we expect coincident point singularities,\footnote{A similar thing happens in non-unitary higher-derivative conformal field theories in certain low dimensions \cite{Brust:2016gjy}, dual to partially massless Vasiliev type theories \cite{Brust:2016zns}.} and we will see that they are not what comes from the AdS/CFT correspondence.   Another way to extract a correlator for the values \eqref{scalarzeoersvalueee} is to look at
the coefficient of the zero by taking $-\Delta\rightarrow k+\epsilon$ in \eqref{pspace2pse}, expanding in $\epsilon$, and reading off the leading non-vanishing term, which is linear in $\epsilon$, 
\be \la \phi(p)\phi(-p)\ra =   \epsilon \,\pi ^{d/2} (-1)^{k+1}   k! 2^{d+2 k}\Gamma \left({d\over 2}+k\right)  p^{-d-2 k} +\cdots\ee
Fourier transforming the expression multiplying this leading $\epsilon$, we get a log in position space,
\be \la \phi(x)\phi(0)\ra =x^{2k}\log(x^2)+\cdots,\label{logcorrelatorxe}\ee
up to ambiguous polynomial terms $\sim x^{2k}$, which set the scale of the log.  

This correlator \eqref{logcorrelatorxe} is not conformally invariant at separated points, because it is not of the form \eqref{position2pse}.  However, the case $k=0$ is familiar in $d=2$: this is the case of a free scalar, which we know has a logarithmic correlator in position space.  In this case, we know that the reason is that $\phi$ itself is not a conformal primary in the theory, rather its derivative $\partial_i\phi$ is a spin $s=1$, $\Delta=1$ primary field that has a standard conformally invariant correlator without logarithms.  We can think of the shift symmetry, $\phi\rightarrow \phi+K$ with constant $K$, as having been gauged, so that we are only supposed to consider operators that are invariant under this symmetry.  The ``field strength'' $\partial_i\phi$ is the basic such operator.

We can interpret the other values of $k$ in \eqref{logcorrelatorxe}, and the other values of $d$, in a similar way.  We can consider \eqref{logcorrelatorxe} to be the correlators of shift symmetric scalars with shift symmetry
\be \phi \rightarrow \phi+K,\label{Kshiftse}\ee
where $K$ is now an order $k$ generalized conformal Killing scalar satisfying
\be \partial_{(i_1}\cdots\partial_{i_{k+1})_T}K=0.\label{conformalkillingse}\ee
(See appendix B of \cite{Brust:2016gjy} for more details about these generalized conformal Killing scalars and tensors.)
Considering \eqref{Kshiftse} as a gauge symmetry, the invariant ``field strength" is a rank $k+1$ symmetric traceless tensor made from $k+1$ derivatives of the scalar,
\be F_{i_1\cdots i_{k+1}} = \partial_{(i_i}\cdots\partial_{i_{k+1})_T}\phi.\label{fieldstrengthsee}\ee
The correlator of \eqref{fieldstrengthsee}, computed from \eqref{logcorrelatorxe} by taking the appropriate derivatives, now has no logs and takes the form \eqref{gencoonfprimarsse} for a primary of spin $s=k+1$ and weight $\Delta=1$,
\be \la F_{i_1\cdots i_{k+1}}(x) F^{j_1\cdots j_{k+1}} (0)\ra= {1\over x^{2}}I_{( i_1}^{( j_1}\cdots I_{ i_{k+1})_T}^{ j_{k+1})_T} . \label{s0fsaqce1}\ee 
In particular, note that the ambiguous polynomial terms $\sim x^{2k}$ which set the scale of the log in \eqref{logcorrelatorxe} vanish in \eqref{fieldstrengthsee}.

Thus we can think of the $\Delta_-$ values of the shift symmetric scalars as theories of a $\Delta=1$ spin $k+1$ primary, the result of a gauged higher shift symmetry.  A $\Delta=1$ spin $k+1$ primary such as this satisfies the following shortening condition as a consequence of the conformal algebra \cite{Penedones:2015aga},
\be\partial_{[i} F_{i_1]i_2\cdots i_{k+1}}=0\,,  \label{spin00binachie}\ee
and indeed this is satisfied by the correlator \eqref{s0fsaqce1}.  This can be thought of as a kind of Bianchi identity, telling us that the field strength \eqref{fieldstrengthsee} is constructed from the ``gauge field'' $\phi$.

The operator \eqref{fieldstrengthsee} carries the $\Delta_-$ value for a spin $k+1$ depth $0$ partially massless field, which is precisely the parent field whose longitudinal mode is the shift symmetric scalar.  The expression \eqref{spin00binachie} is reminiscent of the vanishing of the field strength of this parent PM field.

\subsection{Higher Spins}

We now turn to the general higher spin $s\geq 1$ case.  It is convenient to employ the standard method of packaging symmetric traceless tensor operators with an auxiliary null vector $z^i$ (not to be confused with the Poincar\'e radial coordinate $z$), where the null condition $z^2=0$ enforces tracelessness of the field,
\be \phi(x,z)\equiv \phi_{i_1\cdots i_s}(x)z^{i_1}\cdots z^{i_s}.\ee
The original tensor field is recovered by taking the symmetric traceless part of the coefficient of the $z$'s.  

The correlator \eqref{gencoonfprimarsse} can now be written simply as
\be  \la \phi(x, z)\phi(0, z')\ra=\frac{\left[ z\cdot z' \, x^2-2(x\cdot z)\,  (x\cdot z')\right]^s}{x^{2(\Delta+s)}}.\label{spins2ptzse}\ee
Going to momentum space, this becomes
\bea && \la \phi(p, z)\phi(-p, z')\ra \nn \\ 
&& ={ \pi ^{d/2} \over 2^{2 \Delta-d-s} } {  \Gamma \left({d\over 2}+ s- \Delta \right) \over {\Gamma (s+\Delta )} }  p^{2 \Delta -d} \left(\frac{p\cdot z \, p\cdot z'}{p^2}\right)^s \, _2F_1\left(-s,\Delta
   -1;-\frac{d}{2}-s+\Delta +1;\frac{p^2 \, z\cdot z'}{2 \, p\cdot z\, p\cdot z'}\right) \nn\\
   &&= {\pi ^{d/2}\over 2^{2 \Delta -d} {\Gamma (s+\Delta )}} {p}^{2 \Delta -d} \sum _{n=0}^s {\binom{s}{n} (\Delta -1)_n \Gamma \left(\frac{d}{2}+s-n-\Delta \right)  (z\cdot {z'})^n \left(2\frac{ p\cdot {z}\, p\cdot {z'}}{p^2}\right)^{s-n}}\,, \nn \\ \label{momspace2ptse}
   \eea
where $\left(x\right)_n\equiv {\Gamma\left(x+n\right)\over \Gamma\left( n\right)}=x(x+1)\cdots (x+n-1)$ is the Pochhammer symbol.

It will be convenient to express this in a helicity basis \cite{Arkani-Hamed:2015bza,Isono:2018rrb,Isono:2019wex,Sleight:2019hfp}.  There is a basis of helicity projectors 
\be \Pi^{i_1\cdots i_s,j_1\cdots j_s}_{(h)},\ \ \  h=0,1,2,\ldots,s\ . \ee
They are symmetric and traceless in the $i$'s and in the $j$'s, and symmetric upon interchanging all the $i$'s with all the $j$'s,
and they satisfy orthonormality and completeness relations,
\be \Pi^{i_1\cdots i_s,k_1\cdots k_s}_{(h)} \Pi_{k_1\cdots k_s\ (h')}^{\ \ \ \ \  \ j_1\cdots j_s}=\delta_{hh'} \Pi^{i_1\cdots i_s,j_1\cdots j_s}_{(h)},\ \ \ \sum_{h=0}^s \Pi^{i_1\cdots i_s}_{(h)\ \ \ j_1\cdots j_s}=\delta^{( i_1}_{(i_j}\cdots \delta^{i_s)_T}_{i_s)_T}.\ee
To construct these projectors, first define a rank $h$ projection tensor $P^{i_1\cdots i_h,k_1\cdots k_h}$ as follows,
\be z_{i_1}\cdots z_{i_h}P^{i_1\cdots i_h,k_1\cdots k_h}z'_{j_1}\cdots z'_{j_h}={ h!  \over 2^{h} \left(d-3\over 2\right)_h}(z\ast z\,   z'\ast z')^{h/2} C_h^{\left(d-3\over 2\right)}\left(\frac{z\ast z'}{\sqrt{z\ast z\,   z'\ast z'}}\right)  \,,\ee
where 
\be z\ast z'\equiv z_i \left(\delta^{ij}-{p^ip^j\over p^2}\right) z'_j\ .\ee
The desired helicity projectors are then given by
\be \Pi^{i_1\cdots i_s}  _{(h)\ \ j_1\cdots j_s}=  {1\over p^{2(s-h)}} p^{(i_{h+1}} p_{(j_{h+1}} \cdots p^{i_s} p_{j_s} P^{i_1\cdots i_h)_T}_{\ \ \ \ \ \  j_1\cdots j_h)_T}.\ee
They satisfy conservation conditions:
\be  p_{i_1}\cdots p_{i_{s-h+1}} \Pi^{i_1\cdots i_s,j_1\cdots j_s}_{(h)}=0,\ \ \ \ h=1,2,\ldots,s \ \ \ ,\label{projectorconsce}\ee
\be  Y^T_{[s,h+1]} p^{i_{s+1}}\cdots p^{i_{s+h+1}} \Pi^{i_1\cdots i_s,j_1\cdots j_s}_{(h)}=0,\ \ \ \ h=0,1,2,\ldots,s-1  \ \ \ , \label{projectordualconsce} \ee
where $Y^T_{[s,h+1]}$ projects the $i$ indices onto a 2-row traceless tableau with $i_1,\ldots,i_s$ in the first row. 

The momentum space 2-point function \eqref{momspace2ptse} can now be expressed as
\be \la \phi^{i_1\cdots i_s}(p)\phi^{ j_1\cdots j_s}(-p)\ra= { \pi ^{d/2}\over  2^{2 \Delta -d}  }\frac{ \Gamma \left(\frac{d}{2}-\Delta \right)}{\Gamma (\Delta +s)}  p^{2 \Delta -d} \sum_{h=0}^s c_{h,s} \Pi^{i_1\cdots i_s,j_1\cdots j_s}_{(h)} ,\ \ \ \label{momspace2ptse2} \ee
\be c_{h,s}\equiv (\Delta -1)_h (d-1+h-\Delta )_{s-h} \,. \ee
This form is now convenient for seeing all the special values that occur.

\begin{itemize}

\item At the values of $\Delta$ for which $\nu\equiv \Delta-d/2=0,1,2,3,\ldots$ there are poles. At these values the momentum dependence becomes local (the apparent non-locality in the projectors always cancels out for these values) and the expression \eqref{momspace2ptse2} is not the true Fourier transform of the position space expression \eqref{gencoonfprimarsse}.  The Fourier transform has a logarithmic factor, which can be obtained by taking $\Delta\rightarrow{d\over 2}+ \nu+\epsilon$ and extracting the finite part in the Laurent expansion in $\epsilon$,
\bea  \la \phi^{i_1\cdots i_s}(p)\phi^{ j_1\cdots j_s}(-p)\ra&=&  \frac{\pi ^{d/2} (-1)^{\nu +1}}{\nu !  4^{\nu } \Gamma \left(\frac{d}{2}+\nu+s \right)} \left({1\over \epsilon}  p^{2 \nu }+p^{2 \nu } \log (p^2)\right)\sum_{h=0}^s c_{h,s} \Pi^{i_1\cdots i_s,j_1\cdots j_s}_{(h)}  +\cdots  \nn\\
c_{h,s}&=& ({d\over 2}+ \nu -1)_h ({d\over 2}+h- \nu -1)_{s-h} \,. \label{integerconfinvcoee}\eea
The pole term is local and can be cancelled with a local counterterm, whose finite part also controls the scale of the log.  
As in the scalar case, these momentum space logs do not seem to have anything to do with the shift symmetry or partial masslessness of the corresponding AdS field since they occur in odd $d$ for half integer values of $\Delta$ whose bulk fields do not carry the shift symmetry or any PM symmetries.

\item
At the values $\Delta=t+d-1$, $t=0,1,\ldots,s-1$ the coefficients $c_{h,s}$ for $h\leq t$ all vanish.  Using \eqref{projectorconsce}, we see that the correlator then satisfies a partially massless conservation condition 
\be \partial_{i_1}\cdots \partial_{i_{s-t}}  \la \phi^{i_1\cdots i_s}(x)\phi^{ j_1\cdots j_s}(0)\ra=0\, .\ee
These are the $\Delta_+$ values for a normally quantized partially massless spin-$s$ field of depth $t$, and the dual operators satisfy this partially massless conservation condition \cite{Dolan:2001ih}.   This conservation condition follows from the conformal algebra, and is the Type II shortening condition in the language of \cite{Penedones:2015aga}.

\item At the values $\Delta=1-t$, $t=0,1,\ldots,s-1$ the coefficients $c_{h,s}$ for $h> t$ all vanish.  Using \eqref{projectordualconsce}, we see that the correlator satisfies a dual partially massless conservation condition  
\be Y^T_{[s,t+1]} \partial^{i_{s+1}}\cdots \partial^{i_{s+t+1}} \la \phi^{i_1\cdots i_s}(x)\phi^{ j_1\cdots j_s}(0)\ra=0\, .\ee  
These are the $\Delta_-$ values for an alternately quantized partially massless spin-$s$ field of depth $t$.  This conservation condition indicates the vanishing of the gauge invariant PM field strength. It follows from the conformal algebra, and is the Type IV shortening condition in the language of \cite{Penedones:2015aga}. 

\item  At the values of $\Delta$ for which $ k\equiv -\Delta-s=0,1,2,3,\ldots$ there are zeros.  These are the values where the position space correlator is polynomial in $x^i$.  At these values the position space correlator satisfies the conservation condition
\be \partial^{(i_{s+1}} \cdots \partial^{i_{s+k+1}}\la \phi^{i_1\cdots i_s)_T}(x)\phi^{j_1\cdots j_s}(0)\ra=0.\ee
Thus follows from the conformal algebra, and is the Type I shortening condition in the language of \cite{Penedones:2015aga}.

\end{itemize}

The values $ k= -\Delta-s=0,1,2,3,\ldots$ where there are zeros in the expression \eqref{momspace2ptse2} are the values of interest to us since they are the $\Delta_-$ values for the shift symmetric fields.   As in the scalar case, at these values the momentum space expression \eqref{momspace2ptse2} has zeros because the position space correlators \eqref{gencoonfprimarsse} are analytic in $x^i$, and hence do not have the typical coincident space divergence expected in a field theory (and as we will see is not what comes out of AdS/CFT).  As in the scalar case, we can find an alternative by looking at the coefficient of the zero in the momentum space expression.  Taking $-\Delta\rightarrow s+k+\epsilon$ in \eqref{momspace2ptse2}, expanding in $\epsilon$, and reading off the leading non-vanishing term linear in $\epsilon$, 
\bea
\la \phi^{i_1\cdots i_s}(p)\phi^{ j_1\cdots j_s}(-p)\ra &=&  \epsilon \,\pi ^{d/2} (-1)^{k+1}   k! 2^{d+2 k+2s}\Gamma \left({d\over 2}+k+s\right)  p^{-d-2 k-2s} \sum_{h=0}^s c_{h,s} \Pi^{i_1\cdots i_s,j_1\cdots j_s}_{(h)}  +\cdots  \nn\\
c_{h,s}&=&  (-s-k -1)_h (d-1+h+s+k )_{s-h}\,. \eea
Fourier transforming the expression multiplying $\epsilon$ in this leading term, we obtain a 2-point function for a spin-$s$ field with position space logs for $k=-s-\Delta=0,1,2,\ldots$,
\be \la \phi_{i_1\cdots i_s}(x)\phi^{ j_1\cdots j_s}(0)\ra=\log (x^2)x^{2(k+s)}I_{( i_1}^{( j_1}\cdots I_{ i_s)_T}^{ j_s)_T}+\cdots\ \ ,\label{spins2ptzsloge}\ee
up to ambiguities analytic in $x^i$ which set the scale of the log.
The correlator \eqref{spins2ptzsloge} is not conformally invariant, but anticipating that it is related to the shift symmetric fields, we demand that our CFT have the extended shift symmetry
\be \phi_{i_1\cdots i_s} \rightarrow \phi_{i_1\cdots i_s}+K_{i_1\cdots i_s},\label{Kshiftses}\ee
where $K_{i_1\cdots i_s}$ is an order $k$ generalized conformal Killing tensor satisfying
\be \partial_{(i_1}\cdots\partial_{i_{k+1}}K_{j_1\cdots j_s)_T}=0.\label{genckesee}\ee
Considering this as a gauge symmetry, the invariant ``field strength" is a rank $k+s+1$ symmetric traceless tensor
\be F_{i_1\cdots i_{k+s+1}} = \partial_{(i_i}\cdots\partial_{i_{k+1}}\phi_{i_{k+2}\cdots i_{k+s+1})_T}.\label{spinsfieldstrengtheete}\ee
The correlator of \eqref{spinsfieldstrengtheete}, computed from \eqref{spins2ptzsloge} by taking the appropriate derivatives, is now free of logs and independent of the ambiguous terms analytic in $x^i$, and it takes the conformally invariant form \eqref{gencoonfprimarsse} for a primary of spin $k+s+1$ and weight $\Delta=1-s$,
\be \la F_{i_1\cdots i_{k+s+1}}(x) F^{j_1\cdots j_{k+s+1}} (0)\ra= {1\over x^{2(1-s)}}I_{( i_1}^{( j_1}\cdots I_{ i_{k+s+1})_T}^{ j_{k+s+1})_T} .\label{s0fsaqce1se} \ee

Thus we can think of the $\Delta_-$ value of the level $k$ shift symmetric spin $s$ tensor as the theory of a $\Delta=1-s$ spin $k+s+1$ primary, the result of a gauged higher shift symmetry \eqref{Kshiftses}.  A $\Delta=1-s$ spin $k+s+1$ primary such as this satisfies the following shortening condition as a result of the conformal algebra \cite{Penedones:2015aga},
\be Y^T_{[k+s+1,s+1]} \partial^{j_{1}}\cdots \partial^{j_{s+1}}  F^{i_1\cdots i_{k+s+1}}\,,   \label{spin0binachie} \ee
and indeed this is satisfied by the correlator \eqref{s0fsaqce1se}.  This can be thought of as a kind of Bianchi identity, telling us that the field strength \eqref{spinsfieldstrengtheete} is constructed from the ``gauge field'' $\phi_{i_1\cdots i_s}$.

The operator \eqref{spinsfieldstrengtheete} carries the $\Delta_-$ value for a spin $k+s+1$ depth $s$ PM field, which is precisely the parent field whose longitudinal mode is the shift symmetric field.  The expression \eqref{spin0binachie} is reminiscent of the vanishing of the field strength of this parent PM field.

\begin{figure}[h!]
\begin{center}
\epsfig{file=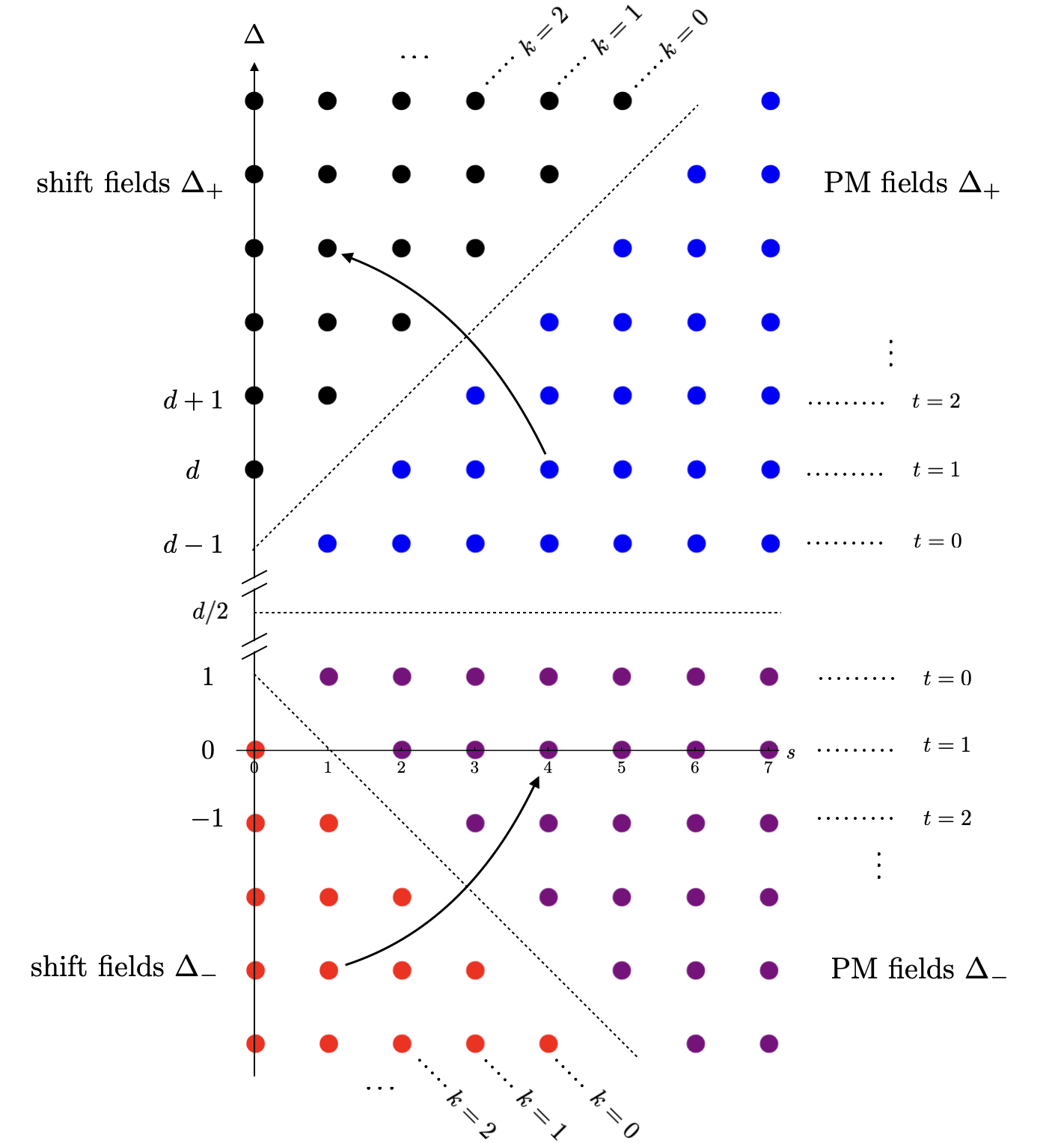,height=5.2in,width=5.0in}
\caption{\small Operator dimensions associated with shift symmetries and PM fields.  The $\Delta_+$ and $\Delta_-$ values are related by reflecting about the line $\Delta=d/2$. The longitudinal mode of a PM field gives the shift field obtained from reflecting about the line $\Delta=s+d-1$, as illustrated by the curved arrow at the top for the $\Delta_+$ values.  An alternately quantized shift field gives a logarithmic position space correlator whose field strength is obtained by reflecting about the line $\Delta=1-s$, which lands on the $\Delta_-$ value for the parent PM field, as illustrated by the curved arrow at the bottom.}
\label{figure1}
\end{center}
\end{figure}

The special values associated with the shift symmetric and PM fields are summarized in Figure \ref{figure1}.  

\section{Shift Symmetries and the Fefferman-Graham Expansion\label{FGsec}}

In the Poincar\'e patch of AdS \eqref{adspoincaremete},
fields have a near-boundary expansion, generally called the Fefferman-Graham expansion, in which the bulk solution is expressed in terms of two pieces of boundary data (see e.g \cite{Bekaert:2012vt,Bekaert:2013zya,Bekaert:2017bpy,Chekmenev:2020lkb}).  In the AdS/CFT correspondence this expansion is used to compute boundary correlators from the bulk theory.  Here we investigate how the shift symmetries act on this expansion, showing that they do not act on one piece of the boundary data, and act on the other by shifting it with a boundary generalized conformal Killing tensor.

The shift symmetries are most conveniently expressed in terms of the ambient embedding space coordinates $X^A$ as in \eqref{eq:tensorshift}.   If we choose light cone coordinates on the ambient space so that $A=(+,-,i)$, $i=1,\ldots,d$ then the ambient metric is
\be  ds^2=-dX^+dX^-+\delta_{ij}dX^i dX^j\, ,\ee
and the AdS$_D$ space is the surface
 \be - X^+X^-+\delta_{ij}X^i X^j=-L^2.\ee
This AdS surface is parametrized in Poincar\'e coordinates by
\be X^+={L^2\over z},\ \ \ X^-={x^2\over z}+z,\ \ \ X^i={L\over z}x^i,\label{poincarecoordseme}\ee
and the induced metric is \eqref{adspoincaremete}.  

Near the boundary $z\rightarrow 0$, we have
\be X^A\rightarrow {L\over z}\left(\tilde X^A+{\cal O}\left(z^2\right)\right)\,,\label{expendnebxe}\ee
where $\tilde X^A(x)$ parametrizes the surface which lies at the intersection of the cone $- X^+X^-+\delta_{ij}X^i X^j=0$ and the plane $X^+=L$,
\be \tilde X^+=L,\ \ \ \tilde X^-={x^2\over L},\ \ \ \tilde X^i=x^i\,.\ee
The induced metric on this surface is $ds^2=\delta_{ij}dx^idx^j$, which can be thought of as the flat metric on the boundary space where the CFT lives.

\subsection{Scalars\label{scalarfgsubsectioner}}

We start with the simplest case of a scalar, $s=0$.  The form of the Fefferman-Graham expansion depends on whether $\nu$, defined as
\be \nu\equiv  \sqrt{{d^2\over 4}+m^2L^2}=\Delta_+-{d\over 2}={d\over 2}-\Delta_-\geq 0, \ \ \ \  s=0\, ,\label{nuscalardef} \ee
is an integer or not.  

\textbf{Non-integer $\nu$:}

For the generic case where $\nu$ is not an integer, the expansion has two different asymptotic falloffs near the boundary and reads
\bea \Phi(z,x)=&& z^{\Delta_-}\left[\phi_{(0)}(x)+z^2\phi_{(2)}(x)+z^4\phi_{(4)}(x)+\cdots\right] \nn \\ 
&& +z^{\Delta_+}\left[\psi_{(0)}(x)+z^2\psi_{(2)}(x)+z^4\psi_{(4)}(x)+\cdots\right],\ \ \ \ \nu\notin{\mathbb Z}.\label{FGexpansionse}\eea
Using the scalar equation of motion \eqref{spinsonsheeqede}, the coefficients $\phi_{(n)}$, $n\geq 2$ are determined in terms of $\phi_{(0)}$, and $\psi_{(n)}$, $n\geq 2$ are determined in terms of $\psi_{(0)}$, so $\phi_{(0)}$ and $\psi_{(0)}$ are the two independent pieces of boundary data.

The shift symmetric scalars have 
\be \Delta_+=d+k,\ \ \ \ \Delta_-=-k\, ,\ \ \  \nu={d\over 2}+k \, ,\ \ \ k=0,1,2,\ldots  \ \ ,\label{shifss2valueese}\ee
and so for odd $d$, $\nu$ is not an integer and the expansion \eqref{FGexpansionse} for a shift symmetric scalar takes the form 
\be  \Phi(z,x)= {1\over z^{k}}\left[\phi_{(0)}(x)+z^2\phi_{(2)}(x)+\cdots\right]  +z^{d+k}\left[\psi_{(0)}(x)+z^2\psi_{(2)}(x)+\cdots\right],\ \ \ \ d\ {\rm odd}. \label{doddscalarfgee}
 \ee 

The shift symmetry acts as
\be \label{eq:scalarambientshift}
\delta \Phi  = S_{A_1\cdots A_k}X^{A_1}\cdots X^{A_k},
\ee
with constant fully symmetric traceless $S_{A_1\cdots A_k}$.
Plugging in \eqref{poincarecoordseme} and expanding in powers of $z$ using \eqref{expendnebxe}, this becomes
\be \label{shiftsymm} \delta \Phi  = \left({L\over z}\right)^k \left[ S_{A_1\cdots A_k}\tilde X^{A_1}\cdots \tilde X^{A_k}+{\cal O}\left(z^2\right)\right]\,.\ee
Matching to the Fefferman-Graham expansion \eqref{doddscalarfgee}, we see that this implies
\be \delta \phi_{(0)}=L^k S_{A_1\cdots A_k}\tilde X^{A_1}\cdots \tilde X^{A_k},\ \ \ \delta \psi_{(0)}=0.\label{deltaphipsieq}\ee
Now, the right hand side of $\delta \phi_{(0)}$ in \eqref{deltaphipsieq} precisely parametrizes the general solution of the generalized conformal Killing scalar equation \eqref{conformalkillingse} \cite{Brust:2016gjy}, so we see that the bulk shift symmetries act as in \eqref{Kshiftse}, i.e. shifts by a generalized conformal Killing scalar, on the boundary field $\phi_{(0)}$.  The other boundary field, $\psi_{(0)}$, is invariant under the shift symmetry.

In standard quantization we take $\phi_{(0)}\sim J$, where $J$ is the source in the generating functional of dual CFT correlators.  We see with this boundary condition, the shift symmetry acts on the source.  This will lead to global Ward identities for the correlation functions in the form of integral constraints.   On the other hand, in the alternate quantization where take $\psi_{(0)}\sim J$, the shift symmetry does not act on the source, and so we expect no Ward identity and thus the shift symmetry to be gauged.   We will confirm these expectations in the next section when we compute correlators from the bulk.

\textbf{Integer $\nu$:}

We now return to the special case where $\nu$ is an integer.  In this case the two different expansions in the Fefferman-Graham expansion overlap and logarithmic terms must be added.  The expansion now reads
\bea \Phi(z,x)=&&z^{\Delta_-}\bigg[\phi_{(0)}(x)+z^2\phi_{(2)}(x)+\cdots+z^{2\nu-2}\phi_{(2\nu-2)}(x)+z^{2\nu}\left(\psi_{(0)}(x)+\phi_{(2\nu)}(x)\log (\mu z)\right)\nn\\
&&+z^{2\nu+2}\left(\psi_{(2)}(x)+\phi_{(2\nu+2)}(x)\log (\mu z)\right)+\cdots\bigg],\ \ \ \nu\in{\mathbb Z} \ .\label{FGexpscalarlongse}\eea
Here $\mu$ is an arbitrary mass scale; its arbitrariness will be related to the arbitrariness of finite contact terms in the dual CFT correlator.
All the coefficients are fixed by the equations of motion in terms of $\phi_{(0)}$ and $\psi_{(0)}$, so these remain as the two independent pieces of boundary data.

From \eqref{shifss2valueese}, we see that these log terms occur in the shift symmetric cases when $d$ is even,
\bea \Phi(z,x)=&&{1\over z^k}\bigg[\phi_{(0)}(x)+z^2\phi_{(2)}(x)+\cdots+z^{d+2k-2}\phi_{(d+2k-2)}(x)+z^{d+2k}\left(\psi_{(0)}(x)+\phi_{(d+2k)}(x)\log (\mu z)\right)\nn\\
&&+z^{d+2k+2}\left(\psi_{(2)}(x)+\phi_{(d+2k+2)}(x)\log (\mu z)\right)+\cdots\bigg],\ \ \ d\ {\rm even} \ .\eea
Looking at \eqref{poincarecoordseme} and \eqref{eq:scalarambientshift}, we see that the highest power of $z$ that appears in the Fefferman-Graham expansion of the shift transformation $\delta\Phi$ is $z^{k}$, so none of the logarithmic terms or $\psi$ terms are affected by the shifts, and we still have \eqref{deltaphipsieq}.

\subsection{Higher Spins}

For a massive higher spin field $s\geq 1$, $\nu$ is now defined by
\be \nu\equiv  \sqrt{{\left(d+2(s-2)\right)^2\over 4}+m^2L^2}=\Delta_+-{d\over 2}={d\over 2}-\Delta_-\geq 0,\ \ \ \ s\geq 1.\label{nudefHS}\ee

The $z$ directed components of the field $\Phi_{z\mu_2\cdots \mu_s}$, and the boundary traces of the boundary components, are determined by constraints in terms of the symmetric traceless boundary components $\Phi_{i_1\cdots i_s}$, so we need only consider these.  

\textbf{Non-Integer $\nu$:}

For non-integer $\nu$, the expansion has two different asymptotic falloffs near the boundary and reads
\bea \Phi_{i_1\cdots i_s}(z,x)=&& z^{\Delta_--s}\left[\phi_{(0) i_1\cdots i_s}(x)+z^2\phi_{(2) i_1\cdots i_s}(x)+z^4\phi_{(4) i_1\cdots i_s}(x)+\cdots\right] \nn \\ 
&& +z^{\Delta_+-s}\left[\psi_{(0) i_1\cdots i_s}(x)+z^2\psi_{(2) i_1\cdots i_s}(x)+z^4\psi_{(4) i_1\cdots i_s}(x)+\cdots\right],\ \ \ \ \nu\notin{\mathbb Z}.\label{FGexpansionseHS}\nn\\\eea
Using the equations of motion \eqref{spinsonsheeqede} and the transversality and tracelessness constraints that eliminate the other fields components, the coefficients $\phi_{(n) i_1\cdots i_s}$, $n\geq 2$ are determined in terms of $\phi_{(0) i_1\cdots i_s}$, and $\psi_{(n) i_1\cdots i_s}$, $n\geq 2$ are determined in terms of $\psi_{(0) i_1\cdots i_s}$, so $\phi_{(0) i_1\cdots i_s}$ and $\psi_{(0) i_1\cdots i_s}$ are the two independent pieces of boundary data.

The shift symmetric fields have
\begin{equation}    \Delta_+=d+k+s\quad\quad \Delta_-=-k-s, \ \ \ \nu={d\over 2}+k+s\, ,\ \ \ k=0,1,2,\ldots \ \ \ ,\label{shifss2valueeses2}
\end{equation}
and so for odd $d$, $\nu$ is not an integer and the expansion takes the form 
\bea \Phi(z,x)_{i_1\cdots i_s}=&& {1\over z^{k+2s}}\left[\phi_{(0) i_1\cdots i_s}(x)+z^2\phi_{(2) i_1\cdots i_s}(x)+z^4\phi_{(4) i_1\cdots i_s}(x)+\cdots\right] \nn \\ 
&& +z^{d+k}\left[\psi_{(0) i_1\cdots i_s}(x)+z^2\psi_{(2) i_1\cdots i_s}(x)+z^4\psi_{(4) i_1\cdots i_s}(x)+\cdots\right],\ \ \ \ d\  {\rm odd}.\nn\\  \label{sdfgexoe} 
\eea

The shift symmetry acts as in \eqref{eq:tensorshift},
\be \label{eq:tensorshift2}
\delta \Phi_{i_1\cdots i_s}= S_{A_1\cdots A_{s+k},B_1\cdots B_s}X^{A_1}\cdots X^{A_{s+k}} {\partial X^{B_1}\over \partial x^{i_1}}\cdots {\partial X^{B_s}\over \partial x^{i_s}} \, ,
\ee
with $S_{A_1\cdots A_{s+k},B_1\cdots B_s}$ a constant traceless field with index symmetries in a 2-row tableau as in \eqref{repKexpre}.  Plugging in \eqref{poincarecoordseme} and expanding in powers of $z$ using \eqref{expendnebxe}, this becomes
\be \label{eq:tensorshifts2e}
\delta \Phi_{i_1\cdots i_s}= \left({L\over z}\right)^{k+2s} \left[  S_{A_1\cdots A_{s+k},B_1\cdots B_s}\tilde X^{A_1}\cdots \tilde X^{A_{s+k}} {\partial \tilde X^{B_1}\over \partial x^{i_1}}\cdots {\partial \tilde X^{B_s}\over \partial x^{i_s}}+{\cal O}\left(z^2\right)\right]\,.\ee

Matching this to the Fefferman-Graham expansion \eqref{sdfgexoe}, we see that this implies
\be \delta \phi_{(0)i_1\cdots i_s}=L^{k+2s}S_{A_1\cdots A_{s+k},B_1\cdots B_s}\tilde X^{A_1}\cdots \tilde X^{A_{s+k}} {\partial \tilde X^{B_1}\over \partial x^{i_1}}\cdots {\partial \tilde X^{B_s}\over \partial x^{i_s}},\ \ \ \delta \psi_{(0)i_1\cdots i_s}=0.\label{deltaphipsieqs}\ee
The right hand side of $\delta \phi_{(0)i_1\cdots i_s}$ in \eqref{deltaphipsieqs} precisely parametrizes the general solution of the generalized conformal Killing tensor equation \eqref{genckesee} \cite{Brust:2016gjy}, so we see that the bulk shift symmetries act as in \eqref{Kshiftses}, i.e. shifts by a generalized conformal Killing tensor, on the boundary field $\phi_{(0)i_1\cdots i_s}$.  The boundary field $\psi_{(0)i_1\cdots i_s}$ is invariant under the shifts.

In the standard quantization we take $\phi_{(0)i_1\cdots i_s}\sim J_{i_1\cdots i_s}$, where $J_{i_1\cdots i_s}$ is the source in the generating functional of dual CFT correlators, and the shift symmetry acts on the source.  In the alternate quantization we take $\psi_{(0)i_1\cdots i_s}\sim J_{i_1\cdots i_s}$, and
 the shift symmetry does not act on the source.

\textbf{Integer $\nu$:}

For integer $\nu$, logarithmic terms must be added to the Fefferman-Graham expansion,
\bea \Phi(z,x)_{i_1\cdots i_s}=&&z^{\Delta_--s}\bigg[\phi_{(0) i_1\cdots i_s}(x)+z^2\phi_{(2) i_1\cdots i_s}(x)+\cdots+z^{2\nu-2}\phi_{(2\nu-2) i_1\cdots i_s}(x) \nn\\
&&+z^{2\nu}\left(\psi_{(0) i_1\cdots i_s}(x)+\phi_{(2\nu) i_1\cdots i_s}(x)\log (\mu z)\right)\nn\\
&&+z^{2\nu+2}\left(\psi_{(2) i_1\cdots i_s}(x)+\phi_{(2\nu+2) i_1\cdots i_s}(x)\log (\mu z)\right)+\cdots\bigg],\ \ \ \nu\in{\mathbb Z} \ .\nn\\\eea
The coefficients are all determined in terms of $\phi_{(0) i_1\cdots i_s}$ and $\psi_{(0) i_1\cdots i_s}$, so these are still the two independent pieces of boundary data.

From \eqref{shifss2valueeses2}, we see that these log terms occur in the shift symmetric cases when $d$ is even,
\bea \Phi(z,x)_{i_1\cdots i_s}=&&{1\over z^{k+2s}}\bigg[\phi_{(0) i_1\cdots i_s}(x)+z^2\phi_{(2) i_1\cdots i_s}(x)+\cdots+z^{d+2k+2s-2}\phi_{(d+2k+2s-2) i_1\cdots i_s}(x) \nn\\
&&+z^{d+2k+2s}\left(\psi_{(0) i_1\cdots i_s}(x)+\phi_{(d+2k+2s) i_1\cdots i_s}(x)\log (\mu z)\right)\nn\\
&&+z^{d+2k+2s+2}\left(\psi_{(2) i_1\cdots i_s}(x)+\phi_{(d+2k+2s+2) i_1\cdots i_s}(x)\log (\mu z)\right)+\cdots\bigg],\ \ \ d\ {\rm even} \ .\nn\\ 
\label{fgmasshsfnine}\eea

Looking at \eqref{poincarecoordseme} and \eqref{eq:tensorshift2}, we see that the mixed symmetry of $S$ ensures that there can be no more than $s+k$ instances of $X^-$ in the shift transformation $\delta\Phi$, so the highest power of $z$ that appears in the Fefferman-Graham expansion of $\delta\Phi$ is $z^{k}$.  Therefore none of the logarithmic terms or $\psi$ terms in \eqref{fgmasshsfnine} are affected by the shifts, and we still have \eqref{deltaphipsieqs}.

\section{AdS/CFT Computations}

In this section we will compute some 2-point correlators for shift symmetric fields from AdS/CFT.  We use the original ``differentiate'' dictionary \cite{Witten:1998qj} in which the on-shell bulk action as a function of the boundary data computes the generating function of the dual CFT correlators.  The divergences are treated with the methods of holographic renormalization \cite{Henningson:1998ey,Henningson:1998gx,deHaro:2000vlm,Skenderis:2000in,Bianchi:2001kw,Skenderis:2002wp}.  

There are two ways of setting boundary conditions on the bulk field, the standard quantization which gives a dual CFT operator of dimension $\Delta_+$, and the alternate quantization which gives a dual CFT operator of dimension $\Delta_-$.  In the case of the shift symmetric fields, the standard quantization gives an operator consistent with the unitarity bounds \eqref{unitboundde}, and the alternate quantization gives an operator that violates these unitarity bounds.  We will see that, as expected from the action of the shift symmetries on the Fefferman-Graham expansion in section \ref{FGsec} and the special values of the 2-point functions studied in section \ref{2ptsec}, it is the non-unitary alternate quantization which shows the shift symmetries.

\subsection{Scalars}

We start with the simplest case, the scalar $s=0$.  The regulated Euclidean action for a free scalar of mass $m$ on AdS$_D$ is,
\bea S_\epsilon&=&{1\over 2L^{D-2}}\int_\epsilon d^DX\sqrt{|G|}\left[(\nabla\Phi)^2+m^2\Phi^2\right]\nn\\
&=& {1\over 2 }\int_\epsilon^\infty dz \int d^dx\ {1\over z^{d-1}}\left[(\partial_z\Phi)^2+\left(\partial_i\Phi\right)^2+{m^2L^2\over z^2}\Phi^2\right]\,, \label{regonshellae}
\eea
where in the second line we have gone to the Poincar\'e coordinates \eqref{adspoincaremete}.
 The factor of $L^{D-2}$ is added to make $\Phi$ dimensionless, and the $z$ integral is cut off near the boundary at $z=\epsilon$.  $1/\epsilon$ serves as the UV regulator in the CFT.
 
 The bulk equation of motion is the Klein-Gordon equation \eqref{spinsonsheeqede},
\be\nabla^2\Phi-m^2\Phi=0.\label{bulkskgee}\ee
Fourier transforming in the $d$ dimensional space, the equation in Poincar\'e coordinates becomes
\be z^{d-1}\partial_z\left[{1\over z^{d-1}}\partial_z \Phi\right]-\left(p^2+{m^2L^2\over z^2}\right)\Phi=0\label{zequation}.\ee
This is then used to recursively solve for the higher order terms in the Fefferman-Graham expansion \eqref{FGexpansionse}, giving
\be \phi_{(n)}(p)={p^2\over n(2\Delta_--d+n)}\phi_{(n-2)}(p),\ \ \  \psi_{(n)}(p)={p^2\over n(2\Delta_+-d+n)}\psi_{(n-2)}(p). \label{recurse}\ee

The regulated action \eqref{regonshellae} reduces on shell to a boundary term, since the bulk term vanishes by the equation of motion \eqref{bulkskgee},
\be S_\epsilon=-\left. {1\over 2}\int d^dx\ {1\over z^{d-1}}\Phi \partial_z\Phi\right|_{z=\epsilon}.\ee
Going to Fourier space $\int d^dx\, \phi(x) f(\partial)\psi(x)\rightarrow\int  {d^dp\over (2\pi)^d} f(-ip) \phi(p)\psi(-p)$, this becomes
\be S_\epsilon=-{1\over 2} \left.\int {d^d p\over  (2\pi)^d}\, {1\over z^{d-1}}  \partial_z \Phi(z,p)\, \Phi(z,-p)  \right|_{z=\epsilon}.\label{eqinte}\ee

To this we must add the counterterm action $S_{\rm c.t.}$, which is constructed from all possible local functions of $\Phi$ and the induced metric $\gamma_{ij}={L^2\over \epsilon^2}\delta_{ij}$, evaluated on the $z=\epsilon$ boundary,
\bea S_{\rm c.t.}&=& \left.{1\over L^d}\int d^d x\  \sqrt{-\gamma}\left[a_0\,\Phi^2 +a_2\,L^2 \Phi\square_{(\gamma)}\Phi+a_4\,L^4\Phi\square^2_{(\gamma)}\Phi+\cdots\right]\right|_{z=\epsilon}\,\nn\\
&=&   \left.\int {d^d x}\ \Phi \left[{1\over \epsilon^d}a_0 +{1\over \epsilon^{d-2}} a_2\, \square+{1\over \epsilon^{d-4}} a_4\,\square^2+\cdots\right]\Phi\right|_{z=\epsilon}\, \nn\\
&=&   \left.\int {d^d p\over (2\pi)^d}\ \Phi(z,p) \left[{1\over \epsilon^d}a_0 -{1\over \epsilon^{d-2}} a_2\, p^2+{1\over \epsilon^{d-4}} a_4\,p^4+\cdots\right]\Phi(z,-p)\right|_{z=\epsilon}\, ,\nn\\ \label{countertermactione}
\eea 
where $\square_{(\gamma)}$ is the spatial Laplacian for the induced metric $\gamma_{ij}$ and $\square$ is the spatial Laplacian for the flat metric $\delta_{ij}$. 
The dimensionless coefficients $a_n$ are constants which are chosen to remove infinities as $\epsilon\rightarrow 0$.  They are independent of $\epsilon$ except for a $\log \epsilon$ dependence in $a_{2\nu}$ in the cases with $\nu\in{\mathbb Z}$.

\textbf{Non-integer $\nu$:} 
 
 For non-integer $\nu$ as defined in \eqref{nuscalardef},
the denominators of \eqref{recurse} never vanish, so we can use this to express all the expansion coefficients in term of $\phi_{(0)}$ and $\psi_{(0)}$. 
 Resumming the resulting series, we get the general solution in terms of these 2 pieces of boundary data,
 \bea \Phi(z,p)&=&{\phi_{(0)}(p)\over 2^{\nu-1}\Gamma(\nu)} p^\nu z^{d/2}K_\nu (pz)  \nn\\  
 && +2^\nu \Gamma(\nu+1)p^{-\nu}\left( \psi_{(0)}(p)-{p^{2\nu}\over 4^\nu}{\Gamma(-\nu)\over\Gamma(\nu)}\phi_{(0)}(p)\right) z^{d/2}I_\nu (pz)\, , \ \ \ \nu\not={\mathbb Z}\, .   \label{fullfgexpanioneg1e}\, 
 \eea

Now we can evaluate the regulated action \eqref{eqinte} by 
plugging in the Fefferman-Graham expansion \eqref{fullfgexpanioneg1e}.
Expanding the result for small $\epsilon$ we find the  divergent terms
\be S_\epsilon=\int  {d^dp\over (2\pi)^d} \ {1\over \epsilon^{2\nu}}b_0 |\phi_{(0)}(p)|^2+{1\over \epsilon^{2\nu-2}}b_2 p^2|\phi_{(0)}(p)|^2+\cdots+{\rm finite} \,, \ \ \nu\not={\mathbb Z}\,,
\ee
with some fixed dimensionless numerical coefficients $b_n$.
Note that the divergent terms depend only on $\phi_{(0)}$, and not on $ \psi_{(0)}$.
The divergent terms are all local terms, and they are removable by the counterterms in $S_{\rm c.t}$.  The coefficients $a_n$ in the counterterm action \eqref{countertermactione} are chosen so as to cancel these divergences,
and this choice is unique.
What is left after this renormalization procedure is the finite renormalized action
\be  S_{\rm ren}=\lim_{\epsilon\rightarrow0}\,(S_\epsilon+S_{\rm c.t.})=\int  {d^dp\over (2\pi)^d}  -{\nu} \, \phi_{(0)}(p)\psi_{(0)}(-p)\ ,\ \ \ \ \ \ \ \nu\not={\mathbb Z}  .\label{phipsiactione}\ee

One of the two boundary data will be fixed by relating it to the source in the dual CFT generating functional.  The other is fixed by requiring that \eqref{fullfgexpanioneg1e} have good falloff behavior in the bulk:
deep in the bulk, we have the behavior
\be K_\nu (pz)\sim e^{-pz},\ \ \ I_\nu (pz)\sim e^{pz},\ \ \ \ z\rightarrow \infty.\label{besselbehaviore}\ee
By demanding that the field fall off at infinity, we get the following non-local relation between $\psi_{(0)}$ and $\phi_{(0)}$, 
\bea \psi_{(0)}(p)&=&{\Gamma(-\nu)\over\Gamma(\nu)}{p^{2\nu}\over 4^\nu}\phi_{(0)}(p)\,, \ \ \nu\not={\mathbb Z} \,.
\label{vanishbulkconde}
\eea

If we want the generating functional in the dual CFT with source $J$, in standard quantization, we are supposed to set the boundary condition $\phi_{(0)}=J$.  This, along with vanishing in the bulk \eqref{vanishbulkconde}, are the two conditions that determine the two boundary data $\phi_{(0)},\psi_{(0)}$.  
Enforcing these two conditions in \eqref{phipsiactione} gives us
\be S_{ren}=\int  {d^dp\over (2\pi)^d}\, -{\nu\over 4^{\nu}}{ \Gamma(-\nu)\over \Gamma(\nu)}p^{2\nu} |J(p)|^2,\ \ \ \nu\not={\mathbb Z},\ \ \ee 
which gives the correct conformally invariant $\sim p^{2\nu}=p^{2\Delta_+-d}$ behavior in \eqref{pspace2pse} of the 2-pt function for a field of weight $\Delta_+$. 

{The shift symmetry acting on the source imposes Ward identities taking the form of integral constraints (analogous to those discussed in \cite{Caldarelli:2016nni} for the case of shift-symmetric axions). These can be found starting with the one-point function of the dual operator in the presence of sources, given by varying the effective action with respect to the source,  
\begin{equation}
    \delta S_{\textrm{ren}}=\int d^dx \langle \mathcal{O}\rangle_J \delta J.
\end{equation} 
Taking the variation $\delta J$ to be a shift symmetry as given in equation \eqref{shiftsymm}, we obtain the Ward identity
\begin{equation}
   \int d^dx \langle \mathcal{O}\rangle_J  S_{A_1\cdots A_k}\tilde X^{A_1}\cdots \tilde X^{A_k}=0. \label{genwardidveve}
\end{equation}
Differentiating this with respect to the source and setting the source to 0 then leads to Ward identities for any given $n$-point function which take the form of integral constraints.   For example, consider the 2-point correlation function of the $k=0$ shift-symmetric scalar, where we have a simple shift symmetry $\delta J=c$. In this case, the Ward identity \eqref{genwardidveve} becomes 
\begin{equation}
   \int d^dx \langle \mathcal{O}\rangle_J =0.
\end{equation}  Differentiating this and setting the source to 0, we see that the 2-point correlator must satisfy the Ward identity
\begin{align}
    \int d^dx\langle\mathcal{O}(x)\mathcal{O}(0)\rangle=0.
\end{align}
In momentum space, this becomes a soft limit, which is indeed satisfied by the momentum space correlator\footnote{In the case of the 2 point function, this constraint is fairly weak and is in fact satisfied by any correlator given by a positive power of $p$, which include those of non-shift symmetric scalars. The Ward identities will become more non-trivial and constraining for higher point correlators in interacting theories, as in \cite{Armstrong:2022vgl}.},
\begin{align}
 \int d^dx\langle\mathcal{O}(x)\mathcal{O}(0)\rangle= \lim_{p\to0} \int d^d x e^{i p\cdot  x}\la \mathcal{O}(x)\mathcal{O}(0)\ra \sim \lim_{p\to0}p^d=0.
\end{align}}

If we want the generating function in alternate quantization  \cite{Klebanov:1999tb} (see \cite{Minces:1999eg,Minces:2002wp,Rivelles:2003ge,Minces:2004ch,Minces:2004zr,Elitzur:2005kz,Hartman:2006dy,Porrati:2016lzr} for more on the alternate quantization) we are supposed to choose $\psi_{(0)}=J$.  Using this along with \eqref{vanishbulkconde} in \eqref{phipsiactione} gives us
\be S_{ren}=\int  {d^dp\over (2\pi)^d}\, -{\nu \, 4^{\nu}}{ \Gamma(\nu)\over \Gamma(-\nu)}{1\over p^{2\nu}} |J(p)|^2,\ \ \ \nu\not={\mathbb Z},\ \ee 
which gives the correct conformally invariant $\sim p^{-2\nu}=p^{2\Delta_--d}$ behavior in \eqref{pspace2pse} for a field of weight $\Delta_-$.

In the shift symmetric cases \eqref{shifss2valueese}, for odd $d$ so that $\nu$ is non-integer, this gives in alternate quantization the momentum space correlator that gives position space logarithms indicating a gauged shift symmetry, as discussed in section \ref{sec1scalarsubsec}.

\textbf{Integer $\nu$:} 

When $\nu$ is an integer, the denominator in the $\psi$ recursion relation in \eqref{recurse} vanishes when $n=2\nu$ and we have to use the Fefferman-Graham expansion \eqref{FGexpscalarlongse} with logs that enter when $n=2\nu$ (we also restrict to $\nu>0$ to avoid additional subtleties that arise in the $\nu=0$ case).  Re-summing this expansion gives
 \bea
\Phi(z,p) &=&\frac{1}{2^{\nu-1}
 \Gamma(\nu)}p^\nu \phi_{0} z^{d/2} K_{\nu}(p z)\nn\\
 &&+ 2^\nu \Gamma(\nu+1) p^{-\nu}\left[\psi_{(0)}+\frac{2(-1)^\nu}{4^\nu\nu \Gamma(\nu)^2}\left(\log(p/\mu)+c\right)p^{2\nu}\phi_{(0)}\right]z^{d/2} I_{\nu}(p z)\,.\nn\\
\eea
Here $c\equiv\frac{1}{2}\gamma-\frac{1}{2}\psi(\nu+1)-\log(2)$ is some fixed numerical constant whose value will be unimportant.

Plugging into \eqref{eqinte} and expanding for small $\epsilon$ we find the divergent terms
\be S_\epsilon=\int d^d x \ {1\over \epsilon^{2\Delta_+-d}}b_0 \phi_{(0)}^2+{1\over \epsilon^{2\Delta_+-d-2}}b_2 \phi_{(0)}\square\phi_{(0)}+\cdots+ \log \epsilon\ b_{2\nu} \phi_{(0)}\square^\nu\phi_{(0)} +{\rm finite} \,, \ \ \nu\in{\mathbb Z}\,.
\ee
There is now a log term present, and the choice of counterterms in the action \eqref{countertermactione} used to remove the divergences are unique except in the choice of $a_{2\nu}$ used to remove the logarithmic divergence: there is an ambiguity in this term reflected in the scale of the log.

The resulting finite on-shell action still takes the form \eqref{phipsiactione},
but there is now also an additional local ambiguity $\sim p^{2\nu}|\phi(p)|^2$ due to the finite ambiguity in the choice of the final counterterm.  

Demanding that the field fall off at infinity now gives the relation
\bea 
\psi_{(0)}(p)&=&-\frac{2(-1)^\nu}{4^\nu\nu \Gamma(\nu)^2}\left(\log(p/\mu)+c\right)p^{2\nu}\phi_{(0)}\, , \ \ \nu\in{\mathbb Z} \, .\label{vanishbulkconde2}
\eea

For the normal quantization $\phi_{(0)}= J$, the result of using \eqref{vanishbulkconde2} in \eqref{phipsiactione} to construct the CFT generating functional is
\be  S_{\rm ren}=\lim_{\epsilon\rightarrow0}\,(S_\epsilon+S_{\rm c.t.})=\int  {d^dp\over (2\pi)^d} {(-1)^\nu\over 2^{2\nu-1}\Gamma(\nu)^2} p^{2\nu}\log(p/\mu) |J(p)|^2,\ \ \ \nu\in{\mathbb Z}\, ,\ \ \label{nuinte2}\ee
up to a local ambiguity $\sim p^{2\nu}|J(p)|^2$, which is reflected in the arbitrariness of the scale $\mu$.
This is the correct anomalously conformally invariant \cite{Bzowski:2013sza,Bzowski:2015pba} behavior $\sim p^{2\Delta_+-d}\log(p)$ of the 2-point function for a field of weight $\Delta_+$, as in \eqref{2scalarlogsolutionee}.
Its Fourier transform is proportional to the correct conformally invariant position space correlator \eqref{position2pse}.

For the alternate quantization $\psi_{(0)}= J$, we instead get a result for the 2-pt. function $\sim {1\over p^{2\nu}(\log(p/\mu)+const.)}$.  This is not what we want for the shift symmetric field in even $d$.  Not only is this not conformally invariant, but none of its derivatives are conformally invariant (so we cannot form field strength correlators from it), and it is not independent of the arbitrary renormalization scale $\mu$ at separated points, indicating some kind of renormalization group flow.  We have not yet found a satisfactory interpretation of this case (see \cite{Porrati:2016lzr} for some similar cases).  There may be an alternative prescription, perhaps a different bulk regulator that also acts on the dimension of operators along the lines of \cite{Bzowski:2016kni,Bzowski:2022rlz}, in which it is easy to see the resolution of this issue.

\subsection{Vectors}

We will not attempt to do the higher spin case in general, we will instead illustrate the main features with the $s=1$ case (see \cite{Mueck:1998iz,Marolf:2006nd,Kabat:2012hp} for other approaches to this case). The regulated Euclidean action for a free vector $A_\mu$ of mass $m$ on AdS$_D$ is 
\bea S_\epsilon &=&{1\over L^{D-4}}\int_\epsilon d^DX\sqrt{|G|}\left[\frac{1}{4}F_{\mu\nu}F^{\mu\nu}+{1\over 2}m^2 A^2\right]\nn\\
&=& \int_\epsilon^\infty dz \int d^dx\ {1\over z^{d-3}}\left[{1\over 2 }( \partial_z A_i-\partial_i A_z )^2+{1\over 4}\left(F_{ij}\right)^2+{1\over 2 }{m^2L^2\over z^2}\left(A_z^2+A_i^2\right) \right]\,, \label{Aregonshellae}
\eea
where $F_{\mu\nu}\equiv \nabla_\mu A_\nu-\nabla_\nu A_\mu$ is the standard Maxwell field strength, and the overall factor of $L$ is chosen so that $A_\mu$ has mass dimension one.

The bulk equation of motion is the Proca equation,
\be \nabla_{\nu}F^{\nu\mu}-m^2A^\mu=0.\ee
{ This is equivalent to the the Klein-Gordon equation \eqref{spinsonsheeqede} together with the transversality constraint,
\be
\left(\nabla^2+{1\over L^2}(D-1)-m^2\right)A_{\mu} = 0,\ \ \ 
\nabla_{\mu}A^\mu=0. \label{blkhdcee}
\ee
}

In Poincar\'e coordinates with the $d$ dimensions Fourier transformed, the $z$ and $i$ components of the Klein-Gordon equation and the transversality constraint become, respectively,
\bea &&\partial_z^2 A_z-\frac{(d-3)}{z}\partial_z A_z-\left(p^2+{m^2L^2+d-1\over z^2}\right)A_z={2i\over z}p^iA_i\, ,\label{Azequation1} \\
&&\partial_z^2 A_i-\frac{(d-3)}{z}\partial_z A_i-\left(p^2+{m^2L^2\over z^2}\right)A_i=-{2i\over z}{p_i}A_z\, , \label{Azequation2} \\
&& \partial_z A_z-{d-1\over z}A_z-ip^iA_i=0. 
\label{Azequation3}
\eea

We now apply the Fefferman-Graham expansion to all the components.
Although the $z$ directed component will be determined on the boundary by constraints in terms the $i$ directed components, it is useful as an intermediate step to apply a Fefferman-Graham expansion to $A_z$ as well. The expansions are given by
\bea A_z(z,x)&=& z^{\Delta_-}\left[\phi_{(0)z }(x)+z^2\phi_{(2)z }(x)+z^4\phi_{(4)z }(x)+\cdots\right] \nn \\ 
&& +z^{\Delta_+}\left[\psi_{(0) z}(x)+z^2\psi_{(2)z}(x)+z^4\psi_{(4)z }(x)+\cdots\right],\nn\\
A_i(z,x)&=&z^{\Delta_--1}\left[\phi_{(0) i}(x)+z^2\phi_{(2)i }(x)+z^4\phi_{(4)i}(x)+\cdots\right] \nn \\ 
&& +z^{\Delta_+-1}\left[\psi_{(0) i}(x)+z^2\psi_{(2) i}(x)+z^4\psi_{(4) i}(x)+\cdots\right]
.\eea
 
Using this in \eqref{Azequation2} we find the relation,
\bea 
\phi_{(n)i}(p)&=&{p^2\over n(2\Delta_--d+n)}\left(\phi_{(n-2)i}(p)-2i\frac{p_i}{p^2}\phi_{(n-2)z}(p)\right),\nn\\
\psi_{(n)i}(p)&=&{p^2\over n(2\Delta_+-d+n)}\left(\psi_{(n-2)i}(p)-2i\frac{p_i}{p^2}\psi_{(n-2)z}(p)\right)
. \label{Arecursezi}\eea
From \eqref{Azequation3} we find expressions relating the $z$ directed and the $i$ directed components of the expansion coefficients, \footnote{In the integer case, the $\psi$ expression is modified to \be \psi_{(n)z}(p)=-{i\,p_i\over (\Delta_--n-1)}\psi_{(n)i}(p)-{i\,p_i\over (\Delta_--n-1)^2}\phi_{(2\nu+n)i}(p), \ee where \be\psi_{(0)z}(p)=-{i\,p_i\over (\Delta_--1)}\psi_{(0)i}(p)+i\frac{(-1)^{\nu}2^{1-2\nu}p^{2\nu}}{(\Delta_--1)(\Delta_+-1)\Gamma(\nu)\Gamma(\nu+1)}p_i\phi_{(0)i}.\ee}
\be
\phi_{(n)z}(p)=-{i\,p_i\over (\Delta_+-n-1)}\phi_{(n)i}(p),\ \ \  \psi_{(n)z}(p)=-{i\,p_i\over (\Delta_--n-1)}\psi_{(n)i}(p)\,.\label{ztoi}
\ee

Using this to eliminate the $z$ directed expansion coefficients in \eqref{Arecursezi}, we get recursion relations in terms of only the $i$ directed components,
\bea 
\phi_{(n)i}(p)&=&{p^2\over n(2\Delta_--d+n)}\left(\delta_{ij}-\frac{2}{(\Delta_+-n+1)}
\frac{p_ip_j}{p^2}\right)\phi_{(n-2)j}(p),\nn\\
\psi_{(n)i}(p)&=&{p^2\over n(2\Delta_+-d+n)}\left(\delta_{ij}-\frac{2}{(\Delta_--n+1)}
\frac{p_ip_j}{p^2}\right)\psi_{(n-2)j}(p)\label{Arecurse}
. \eea
{
The relations \eqref{Arecurse} and \eqref{ztoi} then also solve the remaining equation \eqref{Azequation1}.
}

On shell, the regularized action \eqref{Aregonshellae} reduces to a boundary term,
\be S_\epsilon=-\left. {1\over 2}\int d^dx\ {1\over z^{d-3}}A^i  ( \partial_z A_i-\partial_iA_z)\right|_{z=\epsilon},\ee
which in momentum space becomes
\be S_\epsilon=-{1\over 2} \left.\int {d^d p\over  (2\pi)^d}\, {1\over z^{d-3}} A^i(z,-p) \,
\left(\partial_z A_i(z,p)+ip_i A_z(z,p)   \right)\right|_{z=\epsilon}.\label{Aeqinte}\ee

To this we must add a boundary counterterm action made of all possible local functions of the boundary field $A_i$,
\bea S_{\rm c.t.}&=& {1\over L^d}\int d^d x\  \sqrt{-\gamma}\big[a_0\,A_i\gamma^{ij} A_j +\,L^2 A_i\left(a_2\,\gamma^{ij}\square_{(\gamma)}+b_2\, \partial^i_{(\gamma)} \partial^j_{(\gamma)}\right)A_j\nn\\
&&+\,L^4 A_i\left(a_4\,\gamma^{ij}\square^2_{(\gamma)}+b_4\,\square_{(\gamma)} \partial^i_{(\gamma)} \partial^j_{(\gamma)}\right)A_j+\cdots\big]|_{z=\epsilon}\,\nn\\
&=&   \left.\int {d^d x}\ A_i \bigg[{1\over \epsilon^{d-2}}a_0\,\delta^{ij} +{1\over \epsilon^{d-4}} \left(a_2\,\delta^{ij}\square+b_2\, \partial^i \partial^j\right)+{1\over \epsilon^{d-6}} \left(a_4\,\delta^{ij}\square^2+b_4\,\square \partial^i \partial^j\right)+\cdots\bigg]A_j\right|_{z=\epsilon}\, \nn\\
&=&   \int {d^d p\over (2\pi)^d}\ A_i(z,p) \bigg[{1\over \epsilon^{d-2}}a_0\,\delta^{ij} -{1\over \epsilon^{d-4}} \left(a_2\,\delta^{ij} p^2+b_2\,p^i p^j\right)\nn\\
&&+{1\over \epsilon^{d-6}} \left(a_4\,\delta^{ij}p^4+b_4\,p^2 p^i p^j\right)+\cdots\bigg]A_j(z,-p)|_{z=\epsilon}\, , \label{Acountertermactione}
\eea
where the dimensionless coefficients $a_n$ and $b_n$ will be chosen to remove infinities as $\epsilon\rightarrow 0$.  Note that the divergences will not, and hence the counterterms do not, appear in the local conformally invariant combinations \eqref{integerconfinvcoee}, and so we allow for a separate coefficient for each tensor structure.

\textbf{Non-integer $\nu$:} 
 
For non-integer $\nu$ as defined in \eqref{nudefHS},
the denominators of \eqref{Arecurse} never vanish, allowing for the expression of all the expansion coefficients in term of $\phi_{(0)i}$ and $\psi_{(0)i}$. 
 Resumming the resulting series,
 we get the general solution in terms of the boundary data,
  \bea
A_z(z,p)&=&-{i\,p^i\over 2^{\nu-1}\Gamma(\nu)({d\over 2}+\nu-1)} p^\nu\phi_{(0)i}(p) z^{d/2}K_\nu (pz)  \nn\\  
 && - \frac{i\,2^\nu \Gamma(\nu+1)}{{d\over 2}-\nu-1} p^{-\nu}\left[p^i \psi_{(0)i}(p)-{\Gamma(-\nu)\left({d\over 2}-\nu-1\right)\over 4^\nu \Gamma(\nu)({d\over 2}+\nu-1)}p^{2\nu}p^i\phi_{(0)i}(p)\right] z^{d/2}I_\nu (pz),\nn\\ \label{Azfgsol1}  \\
 A_i(z,p)&=& {\phi_{(0)}^j(p)\over 2^{\nu-1}\Gamma(\nu)} p^\nu\bigg(\delta_{ij}-\frac{2\nu}{{d\over 2}+\nu-1}\frac{p_ip_j}{p^2}\bigg) z^{{d\over 2}-1}K_\nu (pz) +{\phi_{(0)}^j(p)\over 2^{\nu-1}\Gamma(\nu)({d\over 2}+\nu-1)}p^{\nu+1}{p_ip_j\over p^2}z^{d/2}K_{\nu+1}(pz) \nn\\  
 && +2^\nu \Gamma(\nu+1)p^{-\nu}\left[ \psi_{(0)i}(p)-{\Gamma(-\nu)\over 4^\nu\Gamma(\nu)}p^{2\nu}\left(\delta_{ij}-\frac{2\nu}{{d\over 2}+\nu-1}\frac{p_ip_j}{p^2}\right)\phi_{(0)}^j(p)\right] z^{{d\over 2}-1}I_\nu (pz) \nn\\ 
  && -\frac{ 2^\nu \Gamma(\nu+1)}{{d\over 2}-\nu-1} p^{-\nu}{p_i\over p}\left[p_j\psi_{(0)}^j(p)- {\Gamma(-\nu) \left({d\over 2}-\nu-1\right)\over 4^\nu \Gamma(\nu)({d\over 2}+\nu-1)}p^{2\nu} p_j\phi_{(0)}^j(p)\right]z^{d/2}I_{\nu+1} (pz)\,, \nn\\
&& \nu\not={\mathbb Z} \, .  \label{Azfgsol2} \, 
 \eea
 
The regulated vector action \eqref{Aeqinte} can then be evaluated by plugging in this solution. Expanding the result for small $\epsilon$ will lead to divergent terms going as
\be S_\epsilon=\int  {d^dp\over (2\pi)^d} \ \phi_{(0)i}(p)\left[ {1\over \epsilon^{2\nu}}c_0 \delta^{ij}+{1\over \epsilon^{2\nu-2}} (c_2\,p^2\delta^{ij}+d_2\,p^i p^j)+\cdots\right]\phi_{(0)j}(-p)+{\rm finite} \,, 
\ee
with some fixed dimensionless numerical coefficients $c_n$ and $d_n$. They can be removed by the counterterms;  the coefficients $a_n$ and $b_n$ in the counterterm action \eqref{Acountertermactione} are uniquely chosen to cancel these divergences leaving the finite action
\be  S_{\rm ren}=\int  {d^dp\over (2\pi)^d}  -{\nu }\,  \phi_{(0)}^i(p)\psi_{(0)i}(-p)\ ,\ \ \ \ \ \ \ \nu\not={\mathbb Z}  .\label{Aphipsiactione}\ee

Demanding the field be regular in the interior, we find using \eqref{besselbehaviore} in \eqref{Azfgsol2} the following non-local relation between $\psi_{(0)i}$ and $\phi_{(0)i}$, 
\bea  \psi_{(0)i}(p)={\Gamma(-\nu)\over4^\nu\Gamma(\nu)}p^{2\nu}\left(\delta_{ij}-\frac{2\nu}{{d\over 2}+\nu-1}\frac{p_ip_j}{p^2}\right)\phi_{(0)}^j(p)\,, \ \ \nu\not={\mathbb Z}.\label{Avanishbulkconde}
\eea
(Note that the coefficient of $I_{\nu+1}(pz)$ in \eqref{Azfgsol2} and the coefficient of $I_\nu (pz)$ in \eqref{Azfgsol2} are each proportional to $p_i$ contracted with the coefficient of $I_{\nu}(pz)$ in \eqref{Azfgsol2}, so we need only demand that the latter vanishes, which gives \eqref{Avanishbulkconde}.)

To obtain the generating functional in the dual CFT with source $J_i$, in standard quantization, we set the boundary condition $\phi_{(0)i}=J_i$, which together with \eqref{Avanishbulkconde} determines $\phi_{(0)i},\psi_{(0)i}$ and gives us
\be S_{\rm ren}=\int  {d^dp\over (2\pi)^d}\, -{\nu\over 4^{\nu}}{ \Gamma(-\nu)\over \Gamma(\nu)}p^{2\nu} \left(\delta^{ij}-\frac{2\nu}{{d\over 2}+\nu-1}\frac{p^ip^j}{p^2}\right)J_i(p)J_j(-p),\ \ \ \nu\not={\mathbb Z},\ \ \ee 
 giving the correct conformally invariant behavior in \eqref{momspace2ptse2} for a spin-1 field of weight $\Delta_+$.

To obtain the generating function in alternate quantization we choose $\psi_{(0)i}=J_i$, which along with \eqref{Avanishbulkconde} in \eqref{Aphipsiactione} gives us
\be S_{\rm ren}=\int  {d^dp\over (2\pi)^d}\, -{\nu \, 4^{\nu}}{ \Gamma(\nu)\over \Gamma(-\nu)}{1\over p^{2\nu}}\left(\delta^{ij}-\frac{2\nu}{{d\over 2}-\nu-1}\frac{p^ip^j}{p^2}\right) J_i(p)J_j(-p),\ \ \ \nu\not={\mathbb Z},\ \ee 
which gives the correct conformally invariant behavior in \eqref{momspace2ptse2} for a spin-1 field of weight $\Delta_-$.
In alternate quantization, in the shift symmetric cases \eqref{shifss2valueeses2} with odd $d$, this gives the momentum space correlator with position space logarithms indicating a gauged shift symmetry. 

\textbf{Integer $\nu$:} 

When $\nu$ is an integer, we have to use the Fefferman-Graham expansions with logs that enter when $n=2\nu$ (we also restrict to $\nu>0$ to avoid additional subtleties that arise in the $\nu=0$ case),
\bea A_z(z,x)&=&z^{\Delta_-}\bigg[\phi_{(0)}(x)+z^2\phi_{(2)}(x)+\cdots+z^{2\nu-2}\phi_{(2\nu-2)}(x) \nn\\
&&+z^{2\nu}\left(\psi_{(0)}(x)+\phi_{(2\nu)}(x)\log (\mu z)\right)\nn\\
&&+z^{2\nu+2}\left(\psi_{(2)}(x)+\phi_{(2\nu+2)}(x)\log (\mu z)\right)+\cdots\bigg],\nn\\
A_i(z,x)&=&z^{\Delta_--1}\bigg[\phi_{(0) i}(x)+z^2\phi_{(2) i}(x)+\cdots+z^{2\nu-2}\phi_{(2\nu-2) i}(x) \nn\\
&&+z^{2\nu}\left(\psi_{(0) i}(x)+\phi_{(2\nu) i}(x)\log (\mu z)\right)\nn\\
&&+z^{2\nu+2}\left(\psi_{(2) i}(x)+\phi_{(2\nu+2) i}(x)\log (\mu z)\right)+\cdots\bigg],\ \ \ \nu\in{\mathbb Z}
.\eea

The relation in \eqref{ztoi} relating the $z$ and $i$ components also changes in the case $\nu\in{\mathbb Z}$.  The full solution in this case is
{\small
\bea
&& A_z(z,p)= \nn\\
&&-{i\,p_i\over 2^{\nu-1}\Gamma(\nu)({d\over 2}+\nu-1)} p^\nu\phi_{(0)i} z^{d/2}K_\nu (pz)  \nn\\ 
 &&- i\,\frac{2^\nu \Gamma(\nu+1)}{({d\over 2}-\nu-1)} p^{-\nu}p_i\bigg[\psi_{(0)}^i + \frac{(-1)^\nu({d\over 2}-\nu-1)}{({d\over 2}+\nu-1)2^{2\nu-1}\nu\Gamma(\nu)^2}p^{2\nu} \left(\log(p/\mu)+c-{1\over {d\over 2}-\nu-1}\right)\phi_{(0)}^i\bigg]z^{d/2} I_{\nu}(p z)\, ,\nn\\
 \label{Azfgsol1log}  \\
 &&A_i(z,p)= \nn\\ 
 &&{\phi_{(0)}^j\over 2^{\nu-1}\Gamma(\nu)} p^\nu\bigg(\delta_{ij}-\frac{2\nu}{({d\over 2}+\nu-1)}\frac{p_ip_j}{p^2}\bigg) z^{d/2-1}K_\nu (pz) +\frac{\phi_{(0)}^j(p)}{2^{\nu-1}\Gamma(\nu)({d\over 2}+\nu-1)}p^{\nu+1}\frac{p_ip_j}{p^2}z^{d/2}K_{\nu+1}(pz) \nn\\
 && +2^\nu\Gamma(\nu+1) p^{-\nu}\Bigg[\psi_{(0)i}-\frac{(-1)^{\nu}}{2^{2\nu-1}\nu\Gamma(\nu)^2}p^{2\nu} \bigg({\,1\over ({d\over 2}+\nu-1)}\frac{p_ip_j}{p^2}\nn\\
&&-(\log(p/\mu)+c) \left(\delta_{ij}-{2\nu\over ({d\over 2}+\nu-1)}\frac{p_ip_j}{p^2}\right)\bigg)\phi_{(0)}^j\Bigg] z^{d/2-1}I_\nu (pz)\nn\\
 &&-\frac{\Gamma(\nu+1)2^\nu}{({d\over 2}-\nu-1)} p^{-\nu+1}\frac{p_ip_j}{p^2}\bigg[\psi_{(0)}^j +\frac{(-1)^\nu({d\over 2}-\nu-1)}{({d\over 2}+\nu-1)2^{2\nu-1}\nu\Gamma(\nu)^2}p^{2\nu} \left(\log(p/\mu)+c-{1\over {d\over 2}-\nu-1}\right)\phi_{(0)}^j\bigg]z^{d/2}I_{\nu+1}(pz)\,, \nn\\
 &&\ \ \nu\in{\mathbb Z}\,,  \label{Azfgsol2log} \, 
 \eea
 }
 where $c\equiv {1\over 2}{\gamma }-{1\over 2}{\psi(1+\nu)}-\log(2)$ (with $\gamma$ Euler's constant and $\psi$ the digamma function) is a constant whose value will be unimportant.

Plugging into the action \eqref{Aeqinte} and expanding for small $\epsilon$ we find the divergent terms
\bea S_\epsilon&=&\int {d^d p\over (2\pi)^d}\ \phi_{(0)}^i(p)  \bigg[ {1\over \epsilon^{2\nu}}c_0 \delta_{ij}+{1\over \epsilon^{2\nu-2}}\left(c_2\,p^2 \delta_{ij}+d_2\,p_i p_j\right)+\cdots \nn\\
&&+\log (\epsilon)\  c_{2\nu}    \left(\delta_{ij}-{2\nu\over {d\over 2}+\nu-1}\frac{p_ip_j}{p^2}\right) \bigg]\phi_{(0)}^j(-p)+{\rm finite} \,, \ \ \nu\in{\mathbb Z}\,,
\eea
where the $c$'s and $d$'s are some numerical coefficients.
 There is a log divergence present, in addition to the power law divergences.  The relative coefficient between the tensor structures in the log divergence is that of the local conformally invariant combination \eqref{integerconfinvcoee}, whereas the relative coefficient in the power law divergences, and in the finite parts, generally is not.
 
 We then choose the coefficients in the counterterm action \eqref{Acountertermactione} to remove the divergences.  The choice to remove the power law divergences is unique.   The choice of $a_{2\nu}$ and $b_{2\nu}$ used to remove the logarithmic divergence is unique up to an overall ambiguity corresponding to the the scale of the log.   In addition to this, a choice of $a_{2\nu}$ and $b_{2\nu}$ can and should be made so that the finite terms appear in the local conformally invariant combination \eqref{integerconfinvcoee}.  After this, the remaining ambiguity in the finite terms is redundant with the ambiguity in the scale of the log.  
The resulting finite on-shell action still takes the form \eqref{Aphipsiactione},
but with an additional local ambiguity of the form \eqref{integerconfinvcoee}.  

Demanding that the solution \eqref{Azfgsol2log} fall off at infinity gives the relation
\bea 
 \psi_{i0}(p)&=& \frac{(-1)^{\nu}}{2^{2\nu-1}\nu\Gamma(\nu)^2}p^{2\nu} \bigg({\,1\over ({d\over 2}+\nu-1)}\frac{p_ip_j}{p^2}-(\log(p/\mu)+c) \left(\delta_{ij}-{2\nu\over ({d\over 2}+\nu-1)}\frac{p_ip_j}{p^2}\right)\bigg)\phi_{(0)}^j\, \nn\\
 &=& -\frac{(-1)^{\nu}}{2^{2\nu-1}\nu\Gamma(\nu)^2}p^{2\nu} \log(p/\mu)\left(\delta_{ij}-{2\nu\over ({d\over 2}+\nu-1)}\frac{p_ip_j}{p^2}\right)\phi_{(0)}^j+{\rm analytic}\,,\ \ \ \nu\in{\mathbb Z}\,. \nn\\
  \label{Avanishbulkconde2}
\eea
(Note that the coefficient of $I_{\nu+1}(pz)$ in \eqref{Azfgsol2log} and the coefficient of $I_\nu (pz)$ in \eqref{Azfgsol2log} are each proportional to $p_i$ contracted with the coefficient of $I_{\nu}(pz)$ in \eqref{Azfgsol2log}, so we need only demand that the latter vanishes, which gives \eqref{Avanishbulkconde2}.)

For the normal quantization $\phi_{(0)i}= J_i$, the result of using \eqref{Avanishbulkconde2} in \eqref{phipsiactione} to construct the CFT generating functional is
\be  S_{\rm ren}=\int  {d^dp\over (2\pi)^d} {(-1)^\nu\over 2^{2\nu-1}\Gamma(\nu)^2} p^{2\nu}\log(p/\mu)\Big(\delta_{ij}-\frac{2\nu}{{d\over 2}+\nu-1}\frac{p_ip_j}{p^2}\Big) J_i(p)J_j(-p),\ \ \ \nu={\mathbb Z}, \label{Anuinte2}\ee
up to the local ambiguity, which is redundant with the arbitrary scale $\mu$.  Note that the terms other than the log in \eqref{Avanishbulkconde2} are analytic in $p^i$ and only contribute to the local part, and these should also be accounted for when choosing the finite counterterms to keep the local part in the conformal combination.  
The generating functional \eqref{Anuinte2} gives the correct anomalously conformally invariant behavior of the 2-point function for a spin-1 field of weight $\Delta_+$, as in \eqref{2scalarlogsolutionee}.
Its Fourier transform is proportional to the correct conformally invariant position space correlator \eqref{momspace2ptse2}.

Just as in the scalar case, for the alternate quantization $\psi_{(0)i}= J_i$, we get a result for the 2-point function with logs appearing in the denominator, violating conformal invariance at separated points, and with no way to fix it by taking derivatives to form a field strength.  This occurs for the shift symmetric fields in even $d$.  

\section{Discussion}

Shift symmetries play an important role due to their presence in systems with spontaneous symmetry breaking.   They can be used as an organizational tool in categorizing EFTs describing the low energy Goldstone modes resulting from such breaking.
In this work, we opened a study into the properties of the CFTs dual to shift symmetric theories in AdS for general dimension, spin, and shift level. We found that the CFT dual obtained by standard quantization {gives 2-point correlation functions in position space that take the canonical form for primary operators with integer conformal dimensions satisfying the unitarity bound, obeying global Ward identities implied by the shift symmetry.} However, at least in odd CFT dimensions, the shift symmetry leaves signatures in the CFT dual obtained by alternate quantization as a logarithmic structure in their 2-point functions in position space, violating the canonical form for primary operators. This leads to the interpretation of the symmetry as being gauged and the proper observables of the theory as being those that are invariant under the shift symmetry. The shift invariant field strength for a spin $s$ level $k$ shift symmetric field can be constructed by taking $k+1$ traceless symmetrized derivatives of the field, and the 2-point correlation function of the field strength with itself takes the canonical form for a primary operator of spin $k+s+1$ and weight $1-s$.

We studied here only free theories on an AdS background, which is enough to compute dual 2-point functions.  However, some of the shift fields have known fully interacting extensions that preserve the shift symmetries in a non-trivial way.  For example the $k=0,1,2$ scalars lead to AdS versions of the pion chiral Lagrangian, DBI, and special galileon, respectively \cite{Bonifacio:2018zex,Bonifacio:2021mrf,Armstrong:2022vgl}.  Another example is the self-interacting spin 1, $k=0$ theory  \cite{DeRham:2018axr,Bonifacio:2019hrj}, which was found in the massless, fixed curvature decoupling limit of various ghost-free interacting theories for a massive graviton \cite{deRham:2010kj} on AdS.  In this decoupling limit, the tensor interactions vanish leaving only the vector self-interactions. 

For these interacting theories, we can expect the correlation functions of higher point interactions to exhibit interesting behavior. 
In flat space, the interesting behavior manifests as higher order soft theorems for scattering amplitudes, and in the case of the exceptional theories, this soft behavior is strong enough to bootstrap the amplitudes \cite{Cheung:2015ota,Elvang:2018dco}. This naturally leads to the question of what the analogous properties of shift symmetric theories in other spacetimes are. In dS, it was recently demonstrated that, at least at the lowest few orders in powers of the fields, the momentum space boundary wavefunction coefficients of the exceptional theories satisfy soft theorems that can be used to bootstrap the theories \cite{Armstrong:2022vgl} (see also \cite{Bittermann:2022nfh} for the flat space version, and \cite{Assassi:2012zq,Hinterbichler:2013dpa, Green:2022slj} for earlier studies of soft limits of inflationary correlators).
It would be interesting to study the higher point correlators of the CFT duals to the interacting exceptional theories in AdS, and to see how the non-linear shift symmetry manifests. There are some AdS flux vacua where the dual operators have integer conformal dimensions \cite{Conlon:2021cjk,Apers:2022tfm,Plauschinn:2022ztd,Apers:2022vfp}. It would be interesting to further explore whether there is a non-linear shift symmetry behind these.

There are some other directions that could be explored with regards to this work. For example, one could consider the more general $p$-forms and mixed symmetry fields on AdS \cite{Hinterbichler:2022vcc}.  Finally, there is still the open question of how to interpret the odd-$d$ cases where a ${\log(p)}$ appears in the denominator of the 2-point correlator for alternate quantization. A further understanding of this would be useful.

\vspace{-5pt}
\paragraph{Acknowledgments:}

The authors would like to thank Matthias Gaberdiel, Bob Knighton and our anonymous referee for helpful conversations.  KH acknowledges support from DOE grant DE-SC0009946 and Simons Foundation Award Number 658908. 

\bibliographystyle{utphys}
\addcontentsline{toc}{section}{References}
\bibliography{ShiftyBehavior_JHEPv3}

\end{document}